\documentclass[a4paper,11pt]{article}
\usepackage{a4wide}
\usepackage{amsmath}
\usepackage{subcaption}
\usepackage{cite}
\usepackage[normalem]{ulem}
\usepackage{amssymb}
\usepackage{epsf}
\usepackage{amsfonts}
\usepackage{mathrsfs}
\usepackage{graphicx}
\usepackage{tikz}
\usepackage{pgfplots}
\usepackage{braket}
\usepackage{ragged2e}
\usepackage{appendix}
\numberwithin{equation}{section}
\usepackage{bbm}
\usepackage{xcolor}
\usepackage[export]{adjustbox}
\usepackage{siunitx,pslatex}
\usepackage{booktabs}
\usepackage{hyperref}
\usepackage{ulem}
\usepackage{caption}
\usepackage{pgfplots}
\usetikzlibrary{decorations.pathmorphing, decorations.markings, arrows.meta, calc}
\usepgfplotslibrary{fillbetween}

\hypersetup{
    colorlinks,
    linkcolor={green!50!black},
    citecolor={magenta},
    urlcolor={blue!80!black}
}
\graphicspath{{Figures/}}
\usepackage{pgfplots}
\pgfplotsset{compat=1.18}
\begin{document}

\title{\bf Holographic Krylov complexity in confining backgrounds: particle, string, and frames}
	\author{\textbf{Jasper Nongmaithem$^{a}$}\thanks{524ph6005@nitrkl.ac.in},~
    \textbf{Deepak Kumar Kar$^{a}$}\thanks{dkar720@gmail.com},~
    \textbf{Ayush Kumar$^{a}$}\thanks{422ph5005@nitrkl.ac.in},~
    \textbf{Shreeya Rajesh Parab$^{a}$}\thanks{425ph2017@nitrkl.ac.in},~
    \\ \textbf{David Dudal$^{b}$}\thanks{david.dudal@kuleuven.be},~\textbf{Subhash Mahapatra$^{a}$}\thanks{mahapatrasub@nitrkl.ac.in}
			\\\\\textit{{\small $^a$ Department of Physics and Astronomy, National Institute of Technology Rourkela, }}\\
			\textit{{\small Rourkela - 769008, India}}\\
			\textit{{\small $^b$ KU Leuven Campus Kortrijk - Kulak, Department of Physics, }}\\
			\textit{{\small Etienne Sabbelaan 53 bus 7657, 8500 Kortrijk, Belgium}}
			}
    
    \date{}
\maketitle

\begin{abstract}
We investigate holographic Krylov complexity in bottom-up holographic QCD models describing the confined phase, using the recently proposed correspondence between the growth rate of Krylov complexity and the proper radial momentum of probes, such as particles and strings, falling in the bulk spacetime. As representative confining backgrounds, we consider two horizonless geometries, namely the dilaton-deformed AdS-soliton and the deformed thermal-AdS spacetime, both arising as solutions of an Einstein-dilaton gravity system. We find that the Krylov complexity associated with a massive particle exhibits qualitatively distinct behavior in these two backgrounds when analyzed in the Einstein frame. In the deformed AdS-soliton geometry, the complexity undergoes periodic oscillations, with both the oscillation period and amplitude depending on the parameters of the model. In contrast, in the deformed thermal-AdS geometry, the complexity grows exponentially at late times and does not display oscillatory behavior. We then investigate string dynamics in the string frame and find that the Krylov complexity exhibits oscillatory behavior in both confining backgrounds. This behavior is traced to the emergence of an additional turning point in the deep infrared of the string-frame geometry. Motivated by this observation, we further analyze massive-particle dynamics in the string frame, where we find qualitatively similar behavior in the two confining backgrounds. Our results demonstrate that the Krylov complexity of confined holographic systems depends sensitively on the choice of probe and frame, while also revealing a closer correspondence between the two confining geometries in the string-frame description.

\end{abstract}

\section{Introduction}
Quantum chromodynamics (QCD), the fundamental theory of strong interactions, exhibits a rich phase structure arising from its highly nonperturbative nature. Among its most remarkable features is color confinement, i.e., the quarks and gluons are permanently bound inside color-neutral hadrons. Although lattice QCD has remained the most reliable nonperturbative tool for investigating strongly interacting matter, obtaining analytic insights into confinement and its associated dynamical properties remains challenging. In this regard, gauge/gravity duality has emerged as a powerful and complementary framework, providing a novel insight into strongly coupled gauge theories through higher-dimensional gravitational systems \cite{Maldacena:1997re, Gubser:1998bc, Witten:1998qj}. Over the past two decades, holographic methods have been extensively employed to investigate various aspects of QCD, including confinement \cite{Witten:1998zw, Klebanov:2000hb}, hadron physics \cite{Sakai:2004cn, Erlich:2005qh, Armoni:2024gqv}, finite-temperature thermodynamics \cite{Herzog:2006ra, Panero:2009tv, DeWolfe:2010he, Dudal:2017max}, transport properties \cite{Policastro:2001yc, Kovtun:2004de, Jena:2025xcf, Jena:2024cqs}, the phase structure of strongly interacting matter \cite{xia2023structural,asadi2023probing}, and chaotic features \cite{Hashimoto:2016wme, Lilani:2025wnd, Shukla:2023pbp, Shukla:2024qlf, Shukla:2024wsu}. For reviews on holographic QCD, see e.g.~\cite{Casalderrey-Solana:2011dxg, Gursoy:2010fj, Rougemont:2023gfz}. 

Within holographic QCD, the confined phase is generally described by two horizonless geometries: the deformed thermal-AdS and the AdS soliton. The deformed thermal-AdS solution is obtained from the corresponding black-hole geometry by taking the zero-horizon-radius limit. It typically contains a nontrivial exponential warp factor that asymptotes to AdS near the boundary while significantly modifying the deep interior geometry. This deformation gives rise to confinement and several other QCD-like features in the dual boundary theory \cite{Karch:2006pv}. The second confining geometry is the AdS soliton, which is obtained through a double Wick rotation of the AdS black hole \cite{Horowitz:1998ha}. Due to the compactification of one spatial direction, the soliton naturally realizes a confining phase and reproduces characteristic features such as the linear quark-antiquark potential \cite{Witten:1998zw}.\footnote{ Although both the deformed thermal-AdS geometry and the AdS soliton exhibit confining behavior, a comparison of their free energies shows that the AdS soliton is energetically favored over the corresponding deformed thermal-AdS background \cite{Horowitz:1998ha, Mohanty:2026hfg}.} 

While investigations of holographic QCD have mostly been based on conventional observables, such as the Wilson loop, the Polyakov loop, and thermodynamic quantities, recent years have witnessed growing interest in understanding different aspects of QCD from the perspective of quantum information. The underlying motivation is that information-theoretic measures can probe the organization of quantum degrees of freedom in ways that are often inaccessible to conventional observables, thereby providing complementary insights into the dynamics of strongly interacting systems. Consequently, a variety of quantum information measures have been investigated in holographic QCD, including entanglement entropy \cite{Ryu:2006bv, Ryu:2006ef}, mutual information \cite{Headrick:2010zt}, entanglement of purification \cite{Takayanagi:2017knl, Nguyen:2017yqw}, reflected entropy \cite{Dutta:2019gen}, complexity \cite{Susskind:2014rva, Stanford:2014jda, Brown:2015bva}, etc. The investigation of these measures in QCD systems has uncovered several interesting features associated with the confined and deconfined phases, including different imprints of these measures on confined/deconfined phases \cite{Klebanov:2007ws, Nishioka:2006gr, Dudal:2018ztm, Dudal:2016joz, Mahapatra:2019uql, Kol:2014nqa, Jain:2020rbb, Jain:2022hxl, Lewkowycz:2012mw, Jokela:2020wgs}, signatures of confinement/deconfinement transitions \cite{Knaute:2017lll, Ali-Akbari:2017vtb, Arefeva:2020uec, Fujita:2020qvp}, changes in the pattern of quantum correlations and $c$-function flow \cite{Jokela:2025cyz}, behavior of information spreading across different phases of strongly coupled gauge theories \cite{Dudal:2018ztm, Banuls:2022iwk, Jokela:2020wgs}, complexity dynamics \cite{Mahapatra:2018gig, Zhang:2017nth, Chang:2024muq}, etc. 

Among the various quantum information quantities, Krylov complexity has attracted a great deal of attention in quantum many-body dynamics over the last few years. It is a measure of operator growth, and essentially characterizes how an initially simple operator evolves into increasingly complex operators under time evolution by expanding it in a recursively generated orthonormal Krylov basis obtained through the Lanczos algorithm \cite{Parker:2018yvk}.\footnote{Although Krylov complexity was originally introduced to characterize operator growth in the Heisenberg picture, it has subsequently been generalized to the Schr\"{o}dinger picture, where it quantifies the spread of quantum states within the Krylov subspace \cite{Balasubramanian:2022tpr}.} It has emerged as a diagnostic of quantum chaos \cite{Rabinovici:2022beu, Chen:2024imd}, and has been widely used to investigate quantum phase transitions \cite{Afrasiar:2022efk, Caputa:2022eye}, nonequilibrium dynamics \cite{Caputa:2025dep}, thermalization \cite{Alishahiha:2024rwm}, information scrambling \cite{Bhattacharya:2024uxx, Gill:2024acg}, confinement \cite{Jiang:2025wpj}, as order parameter for confinement/deconfinement transition \cite{Anegawa:2024wov}, etc. For more detailed discussion on this topic, see the excellent reviews \cite{Nandy:2024evd, Rabinovici:2023yex, Rabinovici:2025otw, Baiguera:2025dkc}, and references therein.\footnote{Several holographic proposals have been put forward for complexity in the last decade. This includes the widely studied Complexity~$=$~Volume (CV) and Complexity~$=$~Action (CA) proposals \cite{Susskind:2014rva, Stanford:2014jda, Brown:2015bva}. These proposals have successfully reproduced expected features of complexity, most notably its linear late-time growth in black hole backgrounds. For works related to Krylov complexity in various holographic settings, see \cite{Jian:2020qpp, Ambrosini:2024sre, Kar:2021nbm, Heller:2024ldz, Bhattacharya:2026ybq}. }

Recently, a holographic proposal has been suggested for spread complexity  \cite{Caputa:2024sux}. In this proposal, the growth rate of complexity is proportional to the radial momentum of a massive particle propagating in an asymptotically AdS spacetime. The motivation behind this proposal is that the radial momentum of the infalling particle captures the growth of operator size in chaotic quantum systems and therefore naturally reflects the spread of quantum information. Since its introduction, this radial-momentum proposal has found applications in a variety of holographic settings. These include refinements of the original prescription \cite{Fan:2024iop, He:2024pox, Chatzis:2026ekd}, studies of top-down conformal quiver geometries \cite{Fatemiabhari:2025poq},  geometries with internal directions \cite{Fatemiabhari:2025usn, Fatemiabhari:2026goj}, anisotropic systems \cite{BitaghsirFadafan:2026lek}, and analyses involving charged and composite probe particles in Supersymmetric Yang-Mills (SYM) theory and related holographic systems \cite{Fatemiabhari:2025cyy, Zoakos:2026obl, Nastase:2026lhz, Chatzis:2026oou}. For other related works in this direction, see \cite{Alfinito:2026vah, Alfinito:2026yex, Fu:2025kkh, Li:2025fqz, Aguilar-Gutierrez:2025kmw, Aguilar-Gutierrez:2025pqp, Das:2024tnw}.

Despite these developments, the role of Krylov complexity in confining holographic QCD remains relatively unexplored. Existing investigations have largely been restricted to top-down confining backgrounds \cite{Fatemiabhari:2025usn, Fatemiabhari:2026goj, Nunez:2026vhw, Chatzis:2026oou}, while corresponding analyses in phenomenologically motivated bottom-up models are still absent. In the top-down models, the Krylov complexity was found to exhibit periodic behavior in the confined phase, a result that was attributed to the finiteness of the Hilbert space. However, in addition to top-down constructions, there exist phenomenological bottom-up holographic models that capture several essential features of real-world QCD more effectively. Since bottom-up holographic QCD provides a more flexible framework capable of reproducing several salient features of strongly coupled QCD, it is worthwhile to investigate whether confinement leaves similar signatures on the behavior of Krylov complexity in these models. Moreover, as mentioned earlier, the AdS-soliton is not the only horizonless geometry whose dual boundary theory exhibits confinement. The deformed thermal-AdS background, characterized by an exponential warp factor, also represents a confined phase. It is therefore of considerable interest to investigate the behavior of Krylov complexity in this alternative background and determine whether it exhibits features similar to those observed in other confining geometries.

Accordingly, the primary objective of this paper is twofold. First, we investigate holographic Krylov complexity in bottom-up holographic models describing the confined phase of QCD. To this end, we consider the simplest Einstein-scalar gravity model constructed using the potential reconstruction method~\cite{Dudal:2017max}. Such models have attracted considerable attention because they admit analytic or semi-analytic solutions while providing a consistent gravitational framework that captures several qualitative features of QCD. Second, we study the Krylov complexity in two distinct asymptotically AdS geometries describing the confined phase, namely the deformed AdS-soliton and deformed thermal-AdS backgrounds, both of which arise naturally within the same Einstein-scalar framework~\cite{Mohanty:2026hfg}. Following the holographic prescription proposed in~\cite{Caputa:2024sux}, we determine the evolution of the momentum of massive particles and strings in these backgrounds and investigate the resulting behavior of the Krylov complexity.

For massive particles, we find qualitatively different behavior in the two backgrounds when the dynamics is described in the Einstein frame. In the deformed AdS-soliton geometry, the Krylov complexity exhibits periodic oscillations, with the oscillation period depending on the parameters of the holographic model. This behavior is consistent with the results obtained in corresponding top-down constructions~\cite{Fatemiabhari:2025usn,Fatemiabhari:2026goj}. In contrast, in the deformed thermal-AdS background, the complexity grows exponentially at late times and does not exhibit periodic oscillations. We compare these results with those obtained previously in top-down models and examine in detail how the scalar deformation affects the dynamics of information spreading.

We next investigate the Krylov complexity associated with strings in the two confining backgrounds. Since strings couple naturally to the string-frame metric, this analysis is performed in the string frame. Interestingly, in contrast to the particle dynamics in the Einstein frame, the string complexity exhibits qualitatively similar behavior in both confining backgrounds, including a characteristic oscillatory structure. This behavior can be traced to the presence of an additional turning point in the string-frame geometry in the deep interior, which is absent in the corresponding Einstein-frame description. Motivated by this observation, we further examine massive-particle dynamics in the string frame. Remarkably, we find that the particle complexity now displays qualitatively similar behavior in the two confined phases, namely the thermal-AdS and AdS-soliton backgrounds. These results highlight the apparent important role played by the choice of frame in determining the information-spreading dynamics and provide a unified picture of the Krylov complexity in the different confined holographic phases.

The structure of the paper is as follows. In Section~\ref{sec:background}, we introduce the Einstein-dilaton model, present the deformed AdS-soliton and thermal-AdS backgrounds, and specify the relation between the Einstein and string frames. In Section~\ref{sec:pSoliton}, we study massive-particle dynamics in the Einstein frame and compute the associated holographic Krylov complexity in both backgrounds. Section~\ref{sec:extendedop} extends the analysis to falling strings in the string frame. In Section~\ref{sec:particleStrframe}, we examine massive-particle dynamics in the string frame, focusing on the emergence of an infrared turning point and oscillatory complexity in the deformed thermal-AdS background. Finally, we conclude this paper with observations in Section~\ref{Conclusion}.

\section{Confining backgrounds and frame conventions}
\label{sec:background}
To compute the Krylov complexity in the QCD confined phase, we work in the soliton and thermal-AdS geometries in the bulk. These geometries can be constructed from the $5$-dimensional Einstein--dilaton action
\begin{equation}
S_{\rm ED}=\frac{1}{16 \pi G_5} \int \mathrm{d^5}x \ \sqrt{-g}  \ \left[R  -\frac{1}{2}\partial_{M}\phi \partial^{M}\phi -V(\phi)\right]\,,
\label{eq:gravityaction}
\end{equation}
where $G_5$ is the $5$-dimensional Newton constant, $\phi$ is the scalar field, and  $V(\phi)$ is the potential of the field $\phi$. This gravity system exhibits the following soliton-like solution with an asymptotic AdS boundary \cite{Mohanty:2026hfg}:
\begin{eqnarray}
\label{eq:gssol}
ds^2 & = & \frac{\ell^2 e^{2A(z)}}{z^2}\biggl[-dt^2 + \frac{dz^2}{g(z)} + g(z) dx_{s}^2 + \sum\limits_{i=1}^{2} dx_i^2 \biggr] \,, \\ \nonumber
A(z) & = & -a z^2\,,\\ \nonumber
g(z) & = & 1 -  C_s \int_0^z d\xi \ e^{-3A(\xi)} \xi^{3} = 1 - C_s \frac{\left(e^{3 a z^2} \left(3 a z^2-1\right)+1\right)}{18 a^2}   \,, \\ \nonumber
\phi(z)  & = & \sqrt{3} z \sqrt{a \left(2 a z^2+3\right)}+\frac{3}{2} \sqrt{\frac{3}{2}}
   \left(\log (3)-2 \log \left(\sqrt{2 a z^2+3}-\sqrt{2} \sqrt{a}
   z\right)\right)\,, \\ \nonumber
V(z) & = & \frac{3 e^{-2 A(z)} }{\ell^2} \left( - g(z)\left(z^2 A''(z)+3 z^2 A'(z)^2-6 z
   A'(z)+4\right) -\left(z A'(z)-1\right) z g'(z)\right)\,,
\end{eqnarray}
where $\ell$ is the AdS length, which we set to unity from here on, $A(z)$ is the form factor, and $g(z)$ is the capping-off function that governs how the spatial circle $x_s$ smoothly caps off. In particular, the coordinate $x_s$ is periodic with period $L_s$, i.e., $0\leq x_s \leq L_s$, while the other spatial coordinates $\{x_1, x_2\}$ are noncompact. The coordinate $z$ is the usual radial coordinate confined between $z \rightarrow 0$ (UV boundary) to $z = z_0$ (IR end-of-space). The parameter $a$ controls the strength of the dilaton field, and in the limit $a\to 0$, the above solution smoothly reduces to the standard AdS-soliton solution. The integration constant $C_s$ is related to the energy of the soliton and is given by
\begin{eqnarray}
C_s  & = & \frac{1}{\int_{0}^{z_{0}} d\xi \ e^{-3A(\xi)} \xi^{3}} =\frac{18 a^2}{e^{3 a z_0^2} \left(3 a z_0^2-1\right)+1} \,.
\label{Cbsoln2D5}
\end{eqnarray}
The point $z=z_0$ is regular and corresponds to the largest root of $g(z)=0$. At $z=z_0$, the geometry is smooth, as all the curvature invariants are finite there. The regularity 
of the spacetime at $z=z_0$ further fixes the period of the coordinate 
$x_s$ to be
\begin{eqnarray}
\label{solperiod1}
L_{s} & = & -\frac{4 \pi}{g'(z_0)} 
= \frac{4 \pi \int_0^{z_0} \, d\xi e^{-3A(\xi)} \xi^{3}}{z_{0}^{3} e^{-3A(z_0)}}\, = \frac{2 \pi  \left(3 a z_0^2+e^{-3 a z_0^2}-1\right)}{9 a^2 z_0^3}\,.
\end{eqnarray}  
Note that there are only two independent parameters in the soliton (or the confined) phase, namely $z_0$ and $a$. In principle, these parameters can be fixed by taking suitable inputs from the dual boundary field theory. For instance, the parameter $a$ can be fixed from the dual confined/deconfined transition temperature. Similarly, in the standard top-down models (such as the soliton backgrounds obtained by putting $D3$ or $D4$ branes on a circle \cite{Witten:1998zw}), the tip of the soliton $z_0$ is usually related to an intrinsic infrared scale of the theory, such as the mass gap or the effective coupling. In principle, one may attempt to adopt a similar viewpoint in the present bottom-up setup and determine $z_0$ through appropriate phenomenological inputs. However, unlike top-down models, where the underlying string-theoretic construction provides additional constraints, the present framework is more phenomenological in nature, and therefore, such an identification is difficult. In this work, we therefore take a more pragmatic route and treat both $z_0$ and $a$ as effective phenomenological parameters. By exploring a broad range of their values, we aim to systematically investigate how the infrared structure of the geometry affects the behavior of Krylov complexity in the dual confined phase. This also allows us to identify qualitative and potentially universal features of the complexity dynamics without committing to a particular microscopic realization of the dual field theory. This approach is further justified by the fact that the dual boundary theory remains in the confined phase for all values of $z_0$ and $a$ considered here. In particular, the Wilson loop exhibits the characteristic area-law behavior of a confining theory, while the Polyakov loop expectation value continues to vanish.

There exist two other solutions to the Einstein-dilaton equations of motion. The first one corresponds to a black hole with a horizon at $z=z_h$. Essentially, the black hole solution can be obtained by a double analytic continuation $(t_b\rightarrow i x_s,~x_b\rightarrow i t_s)$ of the AdS soliton metric (\ref{eq:gssol}). There exists a first-order phase transition between the black hole and soliton solutions, with the transition point controlled by the ratio of the periods of Euclidean time and the compact space-like
cycle \cite{Mohanty:2026hfg}. In particular, the phase transition occurs at $z_0=z_h$. This phase transition corresponds to the confinement/deconfinement transition in the dual boundary field theory, where the soliton phase represents the confined phase. In contrast, the black hole phase corresponds to the deconfined phase. The soliton phase is dual to the confined phase in the sense that the quark-antiquark potential varies linearly with the separation between them, thereby exhibiting an area law for the Wilson loop, which is a characteristic feature of confinement. Similarly, in this dual confined phase, the Polyakov loop expectation value vanishes. 

The other solution corresponds to deformed thermal-AdS, and is given by
\begin{eqnarray}
ds_{AdS}^2 & = & \frac{\ell^2 e^{2A(z)}}{z^2}\biggl[-dt^2 + dz^2 +dx_{s}^2 + \sum\limits_{i=1}^{2} dx_i^2 \biggr]\,.
\label{thermalAdSback}
\end{eqnarray}
It can essentially be obtained by taking the $z_0 \rightarrow \infty$ limit of the AdS-soliton solution, for which $g_s(z)=1$. Because of the exponential $e^{2A(z)}$ factor, with $A(z)=-a z^2$, the geometry is modified nontrivially in the deep interior while remaining asymptotically AdS near the boundary. This deformation gives rise to confinement and several other QCD-like features in the dual boundary theory \cite{Karch:2006pv, Dudal:2017max, Bohra:2019ebj}. Note that this type of deformed thermal-AdS background is also central to soft-wall holographic QCD models. In the following sections, we shall investigate the Krylov complexity in the confined phases described by both the AdS-soliton and the deformed thermal-AdS backgrounds.

In this work, we are interested in analyzing the Krylov complexity associated with massive particles and strings in the confined phase. Since the dynamics of a fundamental string is naturally described in the string frame, it is useful to express the corresponding bulk backgrounds in both the Einstein and string frames.  The metrics in Eqs.~(\ref{eq:gssol}) and (\ref{thermalAdSback}) are written in the Einstein frame, whereas the string dynamics is governed by the string-frame metric. The two frames are related by the standard dilaton-dependent conformal transformation
$(g_s)_{MN}=e^{\sqrt{2/3}\phi}g_{MN}$.
Applying this transformation to Eq.~(\ref{eq:gssol}), the AdS-soliton geometry in the string frame becomes
\begin{eqnarray}
\label{eq:gssolstring}
ds_{s}^2 & = & \frac{\ell^2 e^{2A_s(z)}}{z^2}\biggl[-dt^2 + \frac{dz^2}{ g(z)} + g(z) dx_{s}^2 + \sum\limits_{i=1}^{2} dx_i^2 \biggr] \,,
\end{eqnarray}
where
\begin{equation}
A_s(z)=A(z)+\sqrt{\frac{1}{6}}\phi(z).
\end{equation}
The corresponding string-frame metric for the deformed thermal-AdS background can be obtained through the same transformation and will be used in the subsequent analysis of string dynamics.

\section{Krylov complexity of the probe particle  in the Einstein frame}\label{sec:pSoliton}
\subsection{In the deformed AdS-soliton background}
In this section, we first study the Krylov (spread) complexity associated with a massive probe particle in the confining phase dual to the AdS-soliton background. To this end, we employ the conjectured correspondence between the growth rate of Krylov complexity and the proper momentum of an infalling probe~\cite{Caputa:2024sux}. We begin by considering a point particle undergoing radial motion in the AdS-soliton geometry described by Eq.~\eqref{eq:gssol}. Since a point particle is minimally coupled to the Einstein-frame metric, we analyze its dynamics in the Einstein frame. Assuming that all other spatial coordinates remain fixed along the geodesic trajectory, the dynamics of the particle is governed only by the coordinates $[z(t),~x_s(t)]$. Using Eq.~\eqref{eq:gssol}, the induced line element formulated in the Einstein frame for the infalling particle simplifies to:
\begin{eqnarray}
    ds_{ind}^2 = \frac{e^{2A(z)}}{z^2} \left( -1 + g(z){\dot{x_s}^2}+\frac{\dot{z}^2}{g(z)} \right) dt^2\,.
    \label{eq:metric}
\end{eqnarray}
The action of the falling point particle is given by,
\begin{equation}\label{falling1}
    \mathcal{S}_P = -m \int dt \mathcal{L}_P, \quad \mathcal{L}_P = \sqrt{-g_{ind}},
\end{equation}
where the Lagrangian $\mathcal{L}_P$ is calculated to be,
\begin{equation}
    \mathcal{L}_P = \frac{e^{A(z)}}{z} \sqrt{1 -  g(z){\dot{x_s}^2}-\frac{\dot{z}^2}{g(z)}}.
\end{equation}
For simplicity, we take $m = 1$ from now on. The canonical momentum conjugate to the $z$ coordinate is $P_z = \frac{\partial L}{\partial \dot{z}}$ (where the standard classical Lagrangian is $L = -\mathcal{L}_P$):
\begin{equation}
    P_z = \frac{\ e^{A(z)}}{z g(z)} \frac{\dot{z}}{\sqrt{1 -g(z){\dot{x_s}^2}- \frac{\dot{z}^2}{g(z)}}}.
\end{equation}
In this setup, there are two conserved quantities. The conserved quantity corresponding to the cyclic coordinate $x_s$ is analogous to the angular momentum of the particle due to the periodic nature of the coordinate $x_s$. The expression for the conserved angular momentum is given by,
\begin{equation}
    J_s = \frac{\ e^{A(z)}}{z} \frac{\dot{x_s} g(z)}{\sqrt{1 -g(z){\dot{x_s}^2}- \frac{\dot{z}^2}{g(z)}}}\,.
\end{equation}
Similarly, the Hamiltonian $\mathcal{H}$, obtained through the Legendre transformation  $\mathcal{H} = P_z \dot{z} +J_s \dot{x}_s + \mathcal{L}_P$, is given by  
\begin{equation}
    \mathcal{H} = \frac{ e^{A(z)}}{z \sqrt{1 -g(z)\dot{x}_s^2- \frac{\dot{z}^2}{g(z)}}}\,,
    \label{eq:Hamiltonian}
\end{equation}
and is conserved due to the time-translation symmetry of the system. To proceed further, we take $\dot{x}_s(t=0)=0$ for simplicity.\footnote{Here we are mainly interested in the radial motion of the particle for simplicity. For works related to the angular motion, one is referred to \cite{Chatzis:2026vzl, Chatzis:2026oou}. } As $J_s$ is a conserved quantity, this makes $J_s=0$ at all times. The Hamiltonian then simplifies to
\begin{equation}
    \mathcal{H} = \frac{e^{A(z)}}{z \sqrt{1 - \frac{\dot{z}^2}{g(z)}}}
    \label{eq:finHamiltonian}
\end{equation}
The above expression is similar to the one found in \cite{Fatemiabhari:2025usn}, however, with an extra exponential term in the numerator. Squaring both sides and solving for $\dot{z}$ yields the following equation of motion:
\begin{equation}
    \dot{z}(t) = \frac{1}{\mathcal{H}z} \sqrt{g(z) \left( \mathcal{H}^2 z^2 - e^{-2az^2} \right)}\,,
    \label{eq:EOM}
\end{equation}
where we have used $A(z)=-a z^2$. We immediately observe that the geodesic possesses two turning points, defined by $\dot{z}(t)=0$. One occurs at $z=z_0$, when $g(z_0)=0$, while the other is determined by the condition $\mathcal{H}^2 z^2 - e^{-2az^2}=0$, corresponding to a finite radial position whose value depends on the conserved energy $\mathcal{H}$. Below, we will set the value of $\mathcal{H}$ using the initial conditions. 

We can integrate the equation to find the trajectory of the particle $z(t)$. The integral would take the form
\begin{equation}
    \int dz  \frac{z}{\sqrt{g(z) \left( \mathcal{H}^2 z^2 - e^{-2az^2} \right)}} = \frac{1}{\mathcal{H}}(t-t_0)\,.
    \label{eq:integral}
\end{equation}
To determine the particle trajectory, we set the integration constant $t_0=0$ and impose the initial conditions $z(0)=\epsilon$ and $\dot z(0)=0$, where $\epsilon$ denotes a small UV cutoff near the asymptotic boundary. Physically, these conditions describe a particle released from rest at the boundary that subsequently falls into the bulk along the radial direction. With this choice, the Hamiltonian takes the value $\mathcal{H}=e^{-a \epsilon^2}/\epsilon$ at the boundary, implying that the particle carries a very large energy scale in the UV limit. In the dual boundary theory, this setup corresponds to preparing a highly energetic UV excitation whose subsequent time evolution is holographically encoded in the bulk motion of the particle, governed by Eq.~(\ref{eq:integral}). In the present bottom-up confining model, due to the complexity of the metric expressions, obtaining an exact closed-form analytic solution for the trajectory is technically difficult. We therefore solve the dynamics numerically within the relevant parameter range. Nevertheless, useful analytic insights can still be extracted in the UV ($z\simeq \epsilon$) and IR ($z\simeq z_0$) regions of the geometry, which help us develop intuition for the overall behavior of the motion.

\subsubsection{Asymptotic limits and geodesic solution}
\label{geodsicsoliton}
Here, we study the behavior of the solution in the UV and IR limits. The integral of Eq.~\eqref{eq:integral}, after substituting the expression of $g(z)$, is expanded in these limits to find the trajectory of the infalling particle. Let us first consider the UV expansion. For simplicity, we denote the quantity under the square root in Eq.~\eqref{eq:integral} as $f_s(z)$.
\begin{equation}
    f_s(z)=g(z)\left(\mathcal H^2z^2-e^{2A(z)}\right),
\qquad
\mathcal H=\frac{e^{A(\epsilon)}}{\epsilon},
\end{equation}
the initial conditions \(z(0)=\epsilon\) and \(\dot z(0)=0\) imply \(f_s(\epsilon)=0\). At fixed \(\mathcal H\), the Taylor expansion gives
\begin{equation}
    f_s(z)=f_s'(\epsilon)(z-\epsilon)+O\!\left((z-\epsilon)^2\right).
\end{equation}
Using the turning-point condition, we obtain
\begin{equation}
    f_s'(\epsilon)=2g(\epsilon)e^{-2a\epsilon^2}
\left(\frac{1}{\epsilon}+2a\epsilon\right)
\end{equation}
Consequently, as $z\rightarrow \epsilon^+$, the integrand behaves as
\begin{align}
     \frac{z}{\sqrt{f_s(z)}} 
    &= \frac{J}{\sqrt{z-\epsilon}}+O\!\left((z-\epsilon)^{1/2}\right) .
\end{align}
where 
\begin{equation*}
    J=\frac{\epsilon}{\sqrt{f_s'(\epsilon)}}=\frac{\epsilon^{3/2}e^{a\epsilon^2}}
{\sqrt{2g(\epsilon)(1+2a\epsilon^2)}}.
\end{equation*}
Accordingly, with our representation, Eq.~\eqref{eq:integral} reduces to
\begin{align}
   \frac{1}{\mathcal{\mathcal{H}}}(t-t_0)&=\int ^z d\bar{z}~ J \frac{1}{\sqrt{\bar{z}-\epsilon}}+O\!\left((\bar{z}-\epsilon)^{1/2}\right) = 2 J\sqrt{z-\epsilon} +O\!\left((z-\epsilon)^{3/2}\right)\,.
\end{align}
Using $t_0 = 0$, we can solve the above equation for $z(t)$. The radial trajectory takes the form
\begin{equation}
    z(t) = \epsilon + \frac{1}{4\mathcal{H}^2J^2} t^2 + O(t^4)\,.
\end{equation}
 For fixed background parameters, \(g(\epsilon)=1+O(\epsilon^4)\) near the AdS boundary. Therefore, in the limit $\epsilon \rightarrow 0^+$,
\begin{equation}
    z(t) = \epsilon + \frac{1}{2\epsilon} t^2 + O(t^4) = \sqrt{t^2+\epsilon^2} + O(t^4)\,.
    \label{zsolUV}
\end{equation}
This is an expected result and will be true for any metric that asymptotes to AdS at the boundary, in agreement with the expression obtained in \cite{Caputa:2024sux}. In particular, the geodesic equation of a particle in pure AdS geometry takes the form
\begin{equation}
  z \ddot{z} + \dot{z}^2 -1 = 0\,,
\end{equation}
where the dot indicates the derivative with respect to $t$. Its general time-like solution is
\begin{equation}
z(t)^2 = (t+c_1)^2 + c_2^2 \,,
\end{equation}
where $c_1$ and $c_2$ are constant. These constants can be fixed from the initial conditions to end up with exactly~(\ref{zsolUV}). Accordingly, the geodesic in the deformed thermal-AdS background exhibits the same UV behavior as given in Eq.~(\ref{zsolUV}).

We can similarly examine the geodesic behavior in the IR at $z = z_0$. Expanding in $z$ around $z=z_0$, the integrand in Eq.~(\ref{eq:integral}) becomes:
\begin{equation}
     \frac{z}{\sqrt{g(z) \left( \mathcal{H}^2 z^2 - e^{-2az^2} \right)}} = \frac{z_0 e^{-\frac{1}{2} \left(a z_0^2\right)}}{3 a \sqrt{z_0-z} \sqrt{\frac{2 \mathcal{H}^2 z_0^5 e^{2 a z_0^2}-2 z_0^3}{e^{3 a z_0^2} \left(3 a z_0^2-1\right)+1}}}+O\left(\sqrt{z_0-z}\right)\,.
\end{equation}
Accordingly, the geodesic equation reduces to
\begin{align}
    \frac{1}{\mathcal{\mathcal{H}}}(t_e-t)&=\int ^z d\bar{z} \frac{z_0 e^{-\frac{1}{2} \left(a z_0^2\right)}}{3 a \sqrt{z_0-\bar{z}} \sqrt{\frac{2 \mathcal{H}^2 z_0^5 e^{2 a z_0^2}-2 z_0^3}{e^{3 a z_0^2} \left(3 a z_0^2-1\right)+1}}} + O\left((z_0-z)^{3/2}\right)\\
   &= \frac{2 z_0 \sqrt{z_0-z}}{3 a \sqrt{\frac{2 \mathcal{H}^2 z_0^5 e^{3 a z_0^2}-2 z_0^3 e^{a z_0^2}}{e^{3 a z_0^2} \left(3 a z_0^2-1\right)+1}}}+O\left((z_0-z)^{3/2}\right)\,,
   \label{eq:end of space}
\end{align}
where $t_e$ is the time when the particle reaches the end of space, i.e., $z(t = t_e) = z_0$. The above expression can be inverted to get
\begin{equation}
    z(t) = z_0-\frac{9 \left(a^2 z_0 e^{a z_0^2} \left(H^2 z_0^2 e^{2 a z_0^2}-1\right)\right)}{2 \left(H^2 \left(e^{3 a z_0^2} \left(3 a z_0^2-1\right)+1\right)\right)} (t-t_e)^2+O\left((t-t_e)^4\right)\,.
\end{equation}
The behavior of the trajectory near $z=z_0$ differs drastically from that reported in \cite{Fatemiabhari:2025usn}. This difference is expected, since the soliton geometry considered here shares the same asymptotic AdS structure as that of \cite{Fatemiabhari:2025usn}, while differing significantly in the deep bulk region. 

Having analyzed the UV and IR profile of the trajectory, we now move on to discuss its complete radial profile. Unfortunately, it is difficult to solve and obtain a closed-form solution of  Eq.~\eqref{eq:integral} analytically. However, it is straightforward to solve it numerically. We obtain the trajectory $z(t)$ by numerically integrating to find $t(z)$, generating a grid of $(t, z)$ points, and performing a numerical inversion via interpolation.  

\begin{figure}[ht]
    \centering 
    \includegraphics[width=0.8\linewidth]{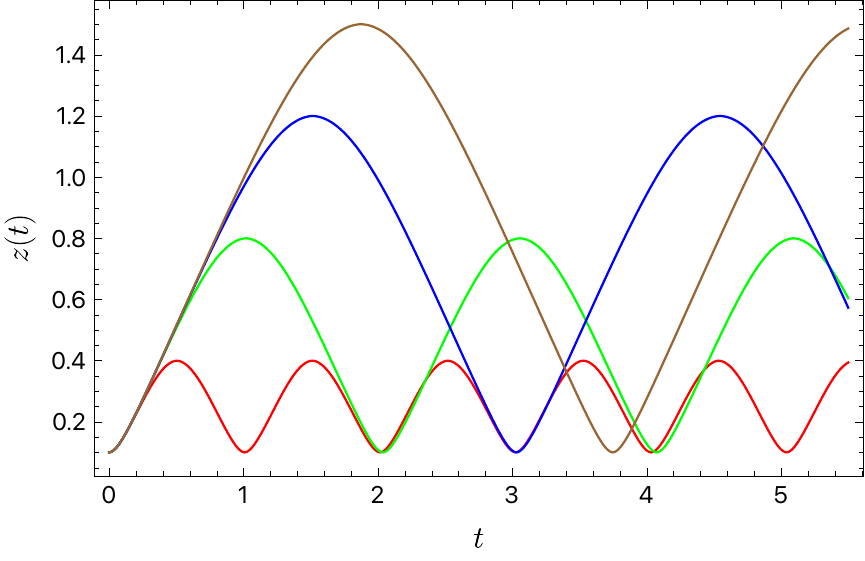}
    \caption{The trajectory $z(t)$ of the probe particle for various values of $z_0$. Here $a = 0.1$ and $\mathcal{H} = 10$ are used. The red, green, blue, and brown curves correspond to $z_0=0.4$, $0.8$, $1.2$, and $1.5$, respectively. }
    \label{fig:trajectoryvsz0}
\end{figure}

\begin{figure}
    \centering
    \includegraphics[width=0.8\linewidth]{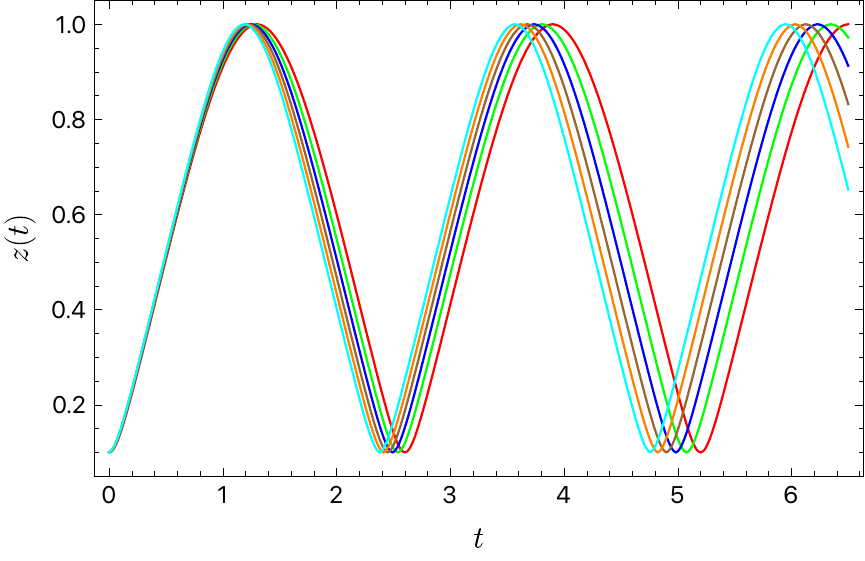}
    \caption{The trajectory $z(t)$ of the probe particle for various values of $a$. Here $z_0 = 1.0$ and $\mathcal{H} = 10$ are used. Red, green, blue, brown, orange, and cyan curves correspond to $a=0$, $0.1$, $0.2$, $0.3$, $0.4$, and $0.5$, respectively. }
    \label{trajectoryvsa}
\end{figure}

Figure~\ref{fig:trajectoryvsz0} shows the time evolution of the radial profile of the particle \(z(t)\). Each colored curve corresponds to a distinct choice of \(z_0\). Because the particle's motion is naturally bounded by the tip of the soliton, its maximum radial depth inherently scales with \(z_0\). However, the nontrivial dynamics are revealed in the oscillation frequencies: for a smaller value of soliton radius (e.g., $z_0=0.4$, red curve), the particle undergoes rapid, high-frequency oscillations. As $z_0$ increases, the motion shifts to lower frequency oscillations. Overall, this demonstrates that while the absolute depth of the particle is tied to the boundary region of the soliton, $z_0$ governs the temporal dynamics and frequency of the motion in the bulk.

In  Figure~\ref{trajectoryvsa}, we keep the IR cutoff fixed at $z_0 = 1.0$ and plot the trajectory with varying $a$ values starting from $0$ to $0.5$. The figure shows a similar oscillatory behavior with the time taken to reach the end of space varying with $a$. Notice that the curves stay together at early times but hit the IR cutoff faster as the strength of the dilaton field increases. This is due to the exponential term in the metric, which dominates later and causes the separation in the trajectories. This can be holographically mapped to the behavior of complexity, showing that the operator spread is faster for larger values of the parameter $a$. We further observe that while the parameter $a$ influences the oscillation frequency, it leaves the amplitude unchanged.

\subsubsection{Krylov complexity from proper momentum}
\label{KC_n_momentum}
In this section, we discuss the procedure for computing the complexity from the proper momentum and analyze its behavior. Our analysis closely follows the approaches developed in~\cite{Caputa:2024sux}. To this end, it is convenient to introduce a new radial coordinate $\bar{\rho}$ defined such that the line element reduces to
$ds^2=d\bar{\rho}^{\,2}$
when all other coordinates in Eq.~\eqref{eq:gssol} are held fixed, i.e., $\Delta t=0$, $\Delta x_i=0$, and $\Delta x_s=0$. In other words, the coordinate $\bar{\rho}$ measures the proper radial distance between two bulk points on a constant-time hypersurface while keeping the remaining spatial directions fixed. Taking these limits for the line element in Eq.~\eqref{eq:metric}, the corresponding proper distance is given by
\begin{equation}
    ds^2_{P} =  \frac{e^{2A(z)} dz^2}{z^2 g(z)}\,.
\end{equation}
Putting $ds_{P } = d\bar{\rho}$, gives us a relation between the old and new radial coordinates
\begin{equation}
    d\bar{\rho} = \frac{e^{A(z)} dz}{z \sqrt{g(z)}}\,,  
\end{equation}
Using the expression of the Hamiltonian in Eq.~\eqref{eq:finHamiltonian}, we get the time derivative of the spatial coordinate $z$,
\begin{equation}
    \dot{z} = \pm \frac{\sqrt{g(z)(\mathcal{H}^2z^2 - e^{2A(z)})}}{\mathcal{H}z}.
\end{equation}
The proper momentum, which according to \cite{Caputa:2024sux} is dual to the rate of change of complexity, is given by
\begin{equation}
    P_{\bar{\rho}} = P_z \frac{d\dot{z}}{d\dot{\bar{\rho}}},
    \label{eq:z_momentum}
\end{equation}
where the radial momentum in the $z$ coordinate is given by
\begin{equation}
        P_z = \frac{e^{A(z)} \dot{z}}{z g(z) \sqrt{1 - \frac{\dot{z}^2}{g(z)}}} = \frac{\dot{z}}{g(z)} \mathcal{H}.
\end{equation}
Substituting the above expression into ~\eqref{eq:z_momentum}, we can compute the proper momentum $P_{\bar{\rho}}$ as a function of the parameters $z$ and $a$. After some algebraic manipulation, the final expression reduces to
\begin{equation}
    P_{\bar{\rho}} = \pm \sqrt{\mathcal{H}^2z^2 e^{-2A(z)} - 1}.
\end{equation}

\begin{figure}
    \centering
    \includegraphics[width=0.8\linewidth]{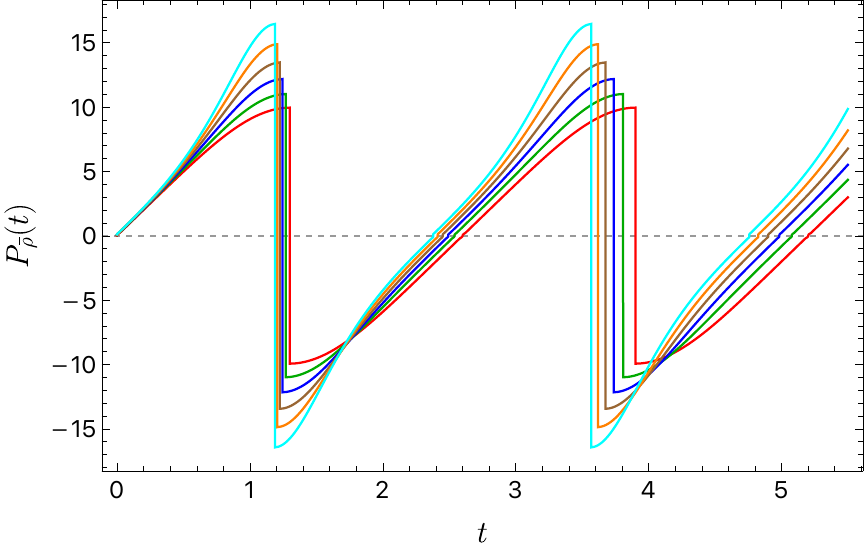}
    \caption{Behavior of proper momentum $P_{\bar{\rho}}(t)$ for different values of $a$. Here, $z_0 = 1$ and $H = 10$ are fixed. Red, green, blue, brown, orange, and cyan curves correspond to $a = 0$, $0.1$, $0.2$, $0.3$, $0.4$, and $0.5$, respectively. } 
    \label{fig:multiple_a_values}
\end{figure}

In Figure~\ref{fig:multiple_a_values}, the behavior of $P_{\bar{\rho}} (t)$ is illustrated for various values of $a$, keeping $z_0=1$ fixed. For any finite value of $a$, an exponential increase in the momentum is observed until the particle reaches the end of the space, at which point it bounces back. In particular, because the soliton geometry ends at a finite radial location $z_0$, a particle initially at rest close to the boundary arrives at the endpoint of the geometry $z_0$ in finite time. The corresponding proper momentum stays finite during the evolution, and changes sign when the particle reaches the endpoint $z_0$ at \(t=t_e\). Interestingly, the time to reach the IR wall decreases with increasing $a$. Accordingly, the time taken to reach the tip of the soliton at $z = z_0$ is faster as $a$ increases. This behavior correlates well with the results shown in Figure~\ref{trajectoryvsa}, where we observed that \(z(t)\) approaches \(z_0\) more rapidly for larger values of \(a\). As time progresses, the oscillatory motion continues as the particle periodically hits the IR boundary.

Now we invoke the relationship suggested in \cite{Caputa:2024sux} to compute the complexity. The rate of change of complexity is defined in terms of the proper momentum as
\begin{equation}
 \partial_t\mathcal{C}(t) = \frac{P_{\bar{\rho}}}{\epsilon}\,,  
\end{equation}
where
\begin{equation}
   P_{\bar{\rho}}= P_z \frac{dz}{d\bar{\rho}} = \pm \sqrt{\mathcal{H}^2z^2 e^{2az^2} - 1}\,.
\end{equation}
This gives us
\begin{equation}
    \mathcal{C}(t) = \int_{0}^{t} d\tilde{t} \frac{P_{\bar{\rho}}\,(\tilde{t})}{\epsilon}\,.
    \label{eq:complexity}
\end{equation}

\begin{figure}[ht]
    \centering
    \includegraphics[width=0.8\linewidth]{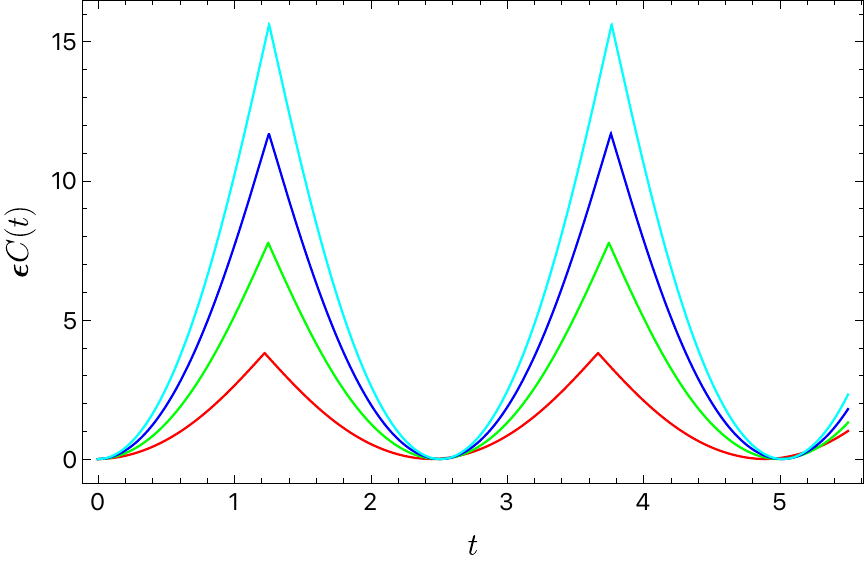}
    \caption{Time dependent growth of the complexity $\epsilon C(t)$ with different $\mathcal{H}$. Here, $a = 0.1$ and $z_0 = 1$ are considered. Red, green, blue, and cyan curves correspond to $\mathcal{H} = 5$, $10$, $15$, and $20$, respectively.}
    \label{fig:hamiltonian}
\end{figure}
\begin{figure}
    \centering
    \includegraphics[width=0.8\linewidth]{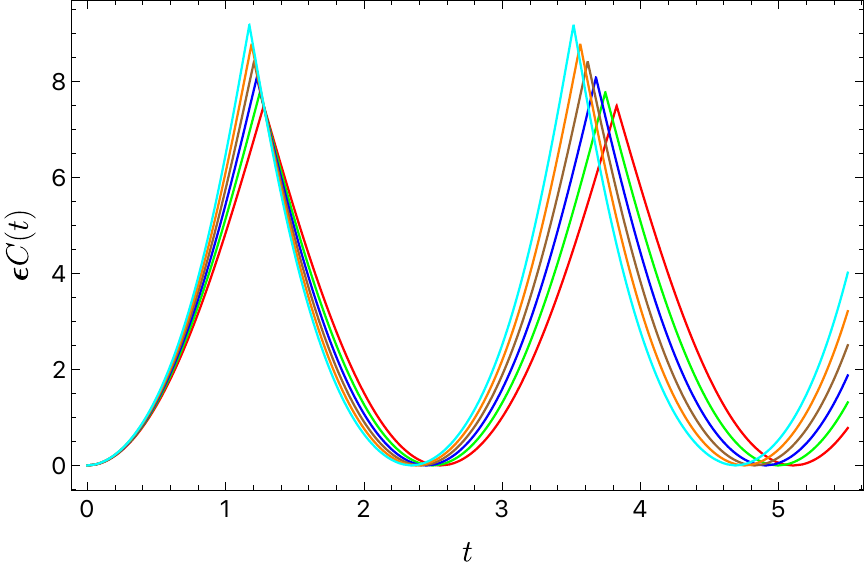}
    \caption{Time dependence of the growth of complexity $\epsilon C(t)$ for different values of $a$. Here, the parameters $\mathcal{H} = 10$ and $z_0 = 1$ are fixed. Red, green, blue, brown, orange, and cyan curves correspond to $a = 0$, $0.1$, $0.2$, $0.3$, $0.4$, and $0.5$, respectively.}
    \label{fig:Complexity_with_a}
\end{figure}

The variation of the holographic complexity calculated from Eq.~\eqref{eq:complexity} with multiple values of $\mathcal{H}$ is presented in Figure~\ref{fig:hamiltonian}. We notice a higher value of complexity for a larger value of the constant $\mathcal{H}$. The higher energy values $\mathcal{H}$ correspond to excitations in the QFT picture. Thus, a larger value of the excitation energy allows the operator complexity to grow much further in the Hilbert space. In the gravitational bulk perspective, a higher energy implies the particle is dropped closer to the asymptotic boundary and thus has a longer time before it hits the IR boundary wall. This extra time allows the radial momentum to grow to a larger value, thereby increasing the complexity in the boundary \cite{Barbon:2019tuq}. In other words, as $t_e$ increases, there is more time for the momentum to increase (this is similar to allowing the operator more time to grow in the Hilbert space). 

Another prominent feature of Figure~\ref{fig:hamiltonian} is the oscillatory behavior. This behavior is observed due to confinement within the upper and lower limits imposed by the UV and IR cutoffs, respectively. Following the speculation in \cite{Nunez:2026vhw}, this might be related to the fact there is a largest and lowest energy in place due to the cut-offs so that only a finite subspace of states can be probed. The operator (dual to the particle) explores all the possible states in a finite time and then recoils back to a more ordered state. This oscillation is a typical response to complexity in the discrete spectrum of two-point correlation functions \cite{Rabinovici:2025otw} and represents a system in which complexity cannot grow indefinitely. From the boundary perspective, this is visualized as the information cannot spread infinitely; it effectively recoils, leading to an oscillating Krylov complexity \cite{Fatemiabhari:2025usn}.

The insight that a stronger strength of the dilaton field causes the operator to spread faster in its Hilbert space, as discussed in Sec.~\ref{geodsicsoliton}, is further solidified in Figure~\ref{fig:Complexity_with_a}. In Figure~\ref{fig:Complexity_with_a}, we plot the Krylov complexity with time by varying the strength of the dilaton field $a$. The red curve represents the standard AdS-soliton geometry, which is recovered in the limit $a \rightarrow 0$. We observe faster growth in complexity and a higher frequency of oscillations for larger values of the parameter $a$. This behavior can be understood from the role of $a$ in controlling the strength of the deformation of the geometry. This deformation causes a change in the proper momentum $P_{\bar{\rho}} = \pm \sqrt{\mathcal{H}^2z^2 e^{2az^2} - 1}$, which in turn causes an increase in the complexity along the particle trajectory. 

Furthermore, increasing $a$ changes the effective radial potential experienced by the infalling particle, causing the turning point to be reached faster. As a result, the particle undergoes repeated bounces between the UV boundary and the IR end of the space over a smaller time interval. From the dual boundary perspective, larger values of the dilaton strength correspond to ``more confinement'' as it increases the deconfinement critical temperature, following the analysis in \cite{Dudal:2017max}, which is visualized by the increased growth rate and rapid oscillatory behavior of the complexity.

\subsection{Krylov complexity in the deformed thermal-AdS background}
\label{krylopAdS}
We now compute the Krylov complexity of the massive probe particle in the deformed thermal-AdS background in the Einstein frame. The background geometry is given in Eq.~(\ref{thermalAdSback}). Since most of the equations and computations are analogous to the AdS-soliton case, here we will be brief and mainly focus on the results. The geodesic equation of the particle now takes the form
\begin{equation}
    \dot{z}(t) = \frac{1}{\mathcal{H}z} \sqrt{\left( \mathcal{H}^2 z^2 - e^{-2az^2} \right)}\,,
    \label{eq:EOMAdS}
\end{equation}
or
\begin{equation}
    \int dz  \frac{z}{\sqrt{\left( \mathcal{H}^2 z^2 - e^{-2az^2} \right)}} = \frac{1}{\mathcal{H}}(t-t_0)\,.
    \label{eq:integralAdS}
\end{equation}
Similar to the AdS-soliton, and unlike the pure AdS case, this integral does not possess a closed-form expression. Nevertheless, much of the relevant physics can still be extracted by considering appropriate limiting regimes. The turning point is given by $\dot{z}=0$. Defining $f(z)=\mathcal{H}^2 z^2 - e^{-2az^2}$, we note that $f'(z)>0$ for any $z>0$, implying the monotonically increasing nature of $f(z)$. Since $f(0)=-1$ and $\lim_{z \to \infty} f(z) = \infty$, it means that for every fixed $\mathcal{H}>0$ and $a>0$, there exists only the UV turning point. Indeed, the turning point equation $\mathcal{H}^2 z_{*}^2 = e^{-2 a z_{*}^2}$ for $\mathcal{H}=\frac{1}{\epsilon}$ translates to $z_*=\epsilon e^{-az_*^2}<\epsilon$, so it cannot be satisfied for any finite $z$ in the deep bulk.

Accordingly, unlike the deformed AdS-soliton case, where there were two turning points, the geodesic does not return back and exhibit oscillatory behavior here. This is similar to the pure thermal-AdS case. 

\begin{figure}[ht]
    \centering
    \includegraphics[width=0.8\linewidth]{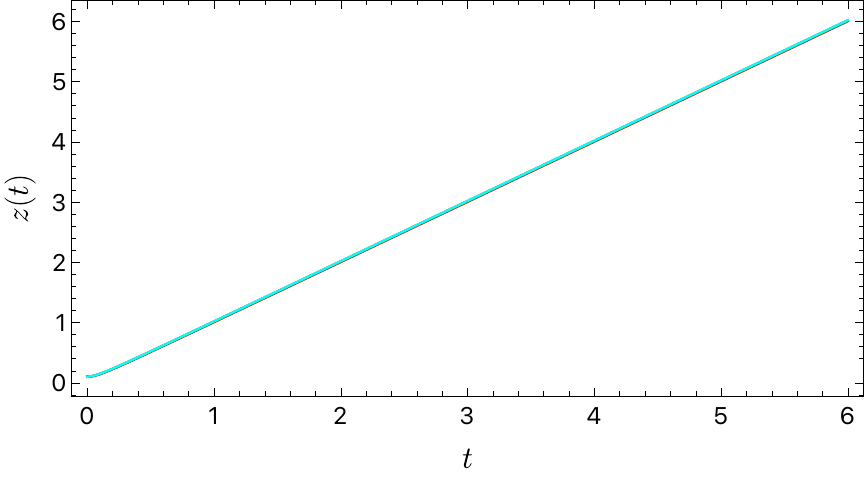}
    \caption{Trajectory $z(t)$ of the probe particle in the deformed thermal-AdS spacetime. Here $\mathcal{H} = 10$ is fixed. Red, green, blue, brown, orange, and cyan curves correspond to $a = 0$, $0.1$, $0.2$, $0.3$, $0.4$, and $0.5$, respectively.}
    \label{fig:TrajectorydeformAdS}
\end{figure}
\begin{figure}
    \centering
    \includegraphics[width=0.8\linewidth]{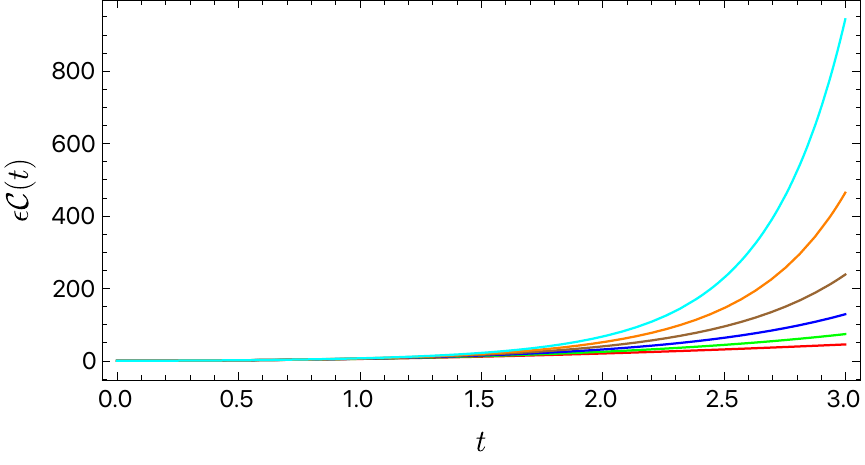}
    \caption{Krylov complexity $\mathcal{C}(t)$ for the deformed thermal AdS space with fixed parameter $\mathcal{H} = 10$. Red, green, blue, brown, orange, and cyan curves correspond to $a = 0$, $0.1$, $0.2$, $0.3$, $0.4$, and $0.5$, respectively.}
    \label{fig:ComplexityDeformedThermal}
\end{figure}

We can similarly evaluate the large $z$ behavior of the geodesic. In this case, the exponential factor $e^{-2 a z^2}$ is negligible. Therefore, we have
\begin{equation}
    \dot{z}(t) = \sqrt{\left( 1 - \frac{e^{-2az^2}}{\mathcal{H}^2 z^2} \right)} = 1 - \frac{e^{-2az^2}}{2 \mathcal{H}^2 z^2}+\dots \,,
\end{equation}
The above equation implies a linear growth of the trajectory,  
\begin{equation}
    \dot{z}(t) \simeq 1\,,~~~~z(t)\sim t\,. 
\end{equation}
which is the same for all $a$. This asymptotic motion is identical to pure AdS and is illustrated in Figure~\ref{fig:TrajectorydeformAdS}. 

The coordinate and proper radial momenta are given by:
\begin{eqnarray}
& & P_z =  \dot{z} \mathcal{H} = \mathcal{H}  \sqrt{1- \frac{e^{-2 a z^2}}{\mathcal{H}^2 z^2}}\,,\qquad P_{\bar{\rho}} =  \mathcal{H} z e^{a z^2} \sqrt{1- \frac{e^{-2 a z^2}}{\mathcal{H}^2 z^2}}\,.
\end{eqnarray}
Near the boundary $z \rightarrow \epsilon$, both the coordinate and proper radial momentum go to zero, as is expected. In the deep interior, as $z \rightarrow \infty$ (late times), since the square root tends to $1$, $P_z$ approaches the constant while the prefactor of $P_{\bar{\rho}}$ goes as $ze^{a z ^2}$. Using the fact that $z(t)\sim t$ at late times, we have the rate of change of complexity
\begin{eqnarray}
& &\partial_t\mathcal{C}(t) \propto t e^{a t^2} \,.
\end{eqnarray}
Or 
\begin{eqnarray}
& &\mathcal{C}(t) \propto e^{a t^2} \,.
\end{eqnarray}
Thus, for all finite values of $a$, the complexity exhibits an exponential growth in the deformed thermal-AdS background (for $a=0$, $\mathcal{C}(t) \propto t^2$). This is certainly very different from the case of the AdS-soliton, where the complexity exhibited oscillatory behavior. 

Let us also point out that since the deformed thermal-AdS geometry considered here is closely related to the background used in soft-wall AdS/QCD models, many of the qualitative results obtained in this section are expected to carry over to the confined phase of soft-wall models as well. 

We end this section by pointing out some similarities and differences between Krylov complexity and entanglement-based measures in the AdS-soliton and deformed thermal-AdS backgrounds. It is well known that entanglement measures, such as the entanglement entropy and the entanglement wedge cross-section, exhibit qualitatively similar behavior in these spacetimes. In both cases, their structure displays a discontinuity as the size of the entangling region is increased. More precisely, there exists a critical size of the entangling region beyond which the connected extremal surface ceases to dominate and is replaced by a disconnected one \cite{Klebanov:2007ws, Jain:2020rbb}. In the AdS-soliton background, this transition originates from the presence of the finite tip at $z=z_0$, which sets the confinement scale and prevents the connected surface from probing arbitrarily deep into the bulk. For the deformed thermal-AdS geometry, although there is no geometric cap analogous to the soliton tip, the exponential warp factor effectively suppresses the penetration of the extremal connected surface into the deep infrared, leading to a qualitatively similar transition \cite{Dudal:2018ztm, Jain:2020rbb}. Consequently, both backgrounds encode confinement through closely related entanglement signatures. 

A similar observation holds for the quark-antiquark potential. In both the AdS-soliton and the deformed thermal-AdS backgrounds, at least in the string frame, the potential exhibits qualitatively similar confining behavior. In the deformed thermal-AdS geometry, the string worldsheet encounters an effective (or ``imaginary'') wall beyond which it cannot propagate. This dynamically generates an infrared confinement scale, resulting in a linear quark-antiquark potential and, equivalently, the area-law behavior of the Wilson loop. In the AdS-soliton background, the finite tip of the geometry plays an analogous role by preventing the string from probing arbitrarily deep into the bulk, leading to the same confining behavior.

The behavior of Krylov complexity, however, is markedly different. While the entanglement measures and quark-antiquark free energy are qualitatively similar in the two confining geometries, the time evolution of Krylov complexity is clearly distinct. In the AdS-soliton background, the falling particle encounters the end of the geometry at the soliton tip and repeatedly bounces back and forth, giving rise to an oscillatory behavior of the complexity. In contrast, the deformed thermal-AdS geometry extends smoothly into the deep bulk without such a reflecting endpoint. As a result, the particle continues to fall towards larger radial distances, leading to a monotonic growth of the complexity. 

\section{Spread complexity for an extended operator}\label{sec:extendedop}
\subsection{In the deformed AdS-soliton background}
In this section, we investigate the spread complexity of an extended (nonlocal) operator in the confining phase dual to the AdS-soliton background. For this purpose, we consider a fundamental string that is extended along one noncompact spatial direction ($x_1$) and falls under gravity in the holographic radial direction. In what follows, we primarily work in the string frame, since a fundamental string couples naturally to the string-frame metric. 

The string motion is described by the Nambu-Goto (NG) action,
\begin{equation}\label{NGaction}
S = -\frac{1}{2\pi\alpha'} \int dt d\sigma \sqrt{-h} \,,
\end{equation}
 where $\alpha'$ is the string tension, $h$ is the determinant of the induced metric $h_{ij} = (g_s)_{\text{MN}}\frac{\partial X^{M}}{\partial \xi^{i}}\frac{\partial X^{N}}{\partial \xi^{j}}$ on the string world sheet, $g_s$ is the metric tensor in the string frame, and $\xi^i$ are the worldsheet coordinates. For the string parameterized by $ \xi^i = (\tau, \sigma)$, a consistent embedding for its motion in the AdS-soliton background would be 
\begin{eqnarray}
t=\tau\,,~~~z=z(t)\,,~~~x_s=x_s(t)\,,~~~x_1=x_1(\sigma)\,.
\label{stringembed}
\end{eqnarray}
On this embedding, the Nambu-Goto (NG) action of the string takes the form
\begin{eqnarray}\label{NGaction}
S & = & -\frac{1}{2\pi\alpha'} \int dt  \sqrt{\left(g_{tt}-g_{zz}\dot{z}^2 - g_{x_s x_s}\dot{x}_s^2 \right)g_{x_1 x_1}} \int d\sigma ~ x_{1}'(\sigma)\,, \nonumber \\
& = & - m \int dt  \sqrt{\left(g_{tt}-g_{zz}\dot{z}^2 - g_{x_s x_s}\dot{x}_s^2 \right)g_{x_1 x_1}} = - m \int dt \mathcal{L}\,, 
\end{eqnarray}
where the dot indicates the derivative with respect to $t$, the prime indicates the derivative with respect to $\sigma$, and we have defined
\begin{eqnarray}
m= \frac{\mathcal{X}_1}{2\pi\alpha'} \,,~~~\text{with}~~   \mathcal{X}_1 = \int d\sigma ~ x_{1}'(\sigma)\,.
\label{mexpstring}
\end{eqnarray}
The canonical momenta associated with $z$ and $x_s$ coordinates are
\begin{eqnarray}
P_z = \frac{m ~ \dot{z} ~ g_{zz} g_{x_1 x_1} }{\sqrt{\left(g_{tt}-g_{zz}\dot{z}^2 - g_{x_s x_s}\dot{x}_s^2 \right)g_{x_1 x_1}}}\,, \nonumber \qquad
J_s = \frac{m ~ \dot{x}_s ~ g_{x_s x_s} g_{x_1 x_1}  }{\sqrt{\left(g_{tt}-g_{zz}\dot{z}^2 - g_{x_s x_s}\dot{x}_s^2 \right)g_{x_1 x_1}}}\,.
\label{conservedmom}
\end{eqnarray}
Since the coordinate $x_s$ is cyclic, the angular momentum $J_s$ is conserved. Similarly, the Hamiltonian $\mathcal{H}$ is given by  
\begin{equation}
    \mathcal{H} = \frac{m g_{tt} g_{x_1 x_1} }{\sqrt{\left(g_{tt}-g_{zz}\dot{z}^2 - g_{x_s x_s}\dot{x}_s^2 \right)g_{x_1 x_1}}}\,,
    \label{eq:Hamiltonian0}
\end{equation}
and is conserved due to the time-translation symmetry of the system. For simplicity, we again take $\dot{x}_s(t=0)=0$ from here on. As $J_s$ is a conserved quantity, this makes $J_s=0$ at all times. The Hamiltonian then simplifies to
\begin{equation}
    \mathcal{H} = \frac{m g_{tt} g_{x_1 x_1} }{\sqrt{\left(g_{tt}-g_{zz}\dot{z}^2 \right)g_{x_1 x_1}}}\,.
    \label{eq:Hamiltonian}
\end{equation}
One can use the Hamiltonian to recast the second-order differential equation into a first-order. Squaring both sides of (\ref{eq:Hamiltonian}) and solving $\dot{z}$, we get the following first-order equation of motion:
\begin{equation}
  \dot{z} = \frac{1}{\mathcal{H}} \sqrt{\frac{g_{tt}}{g_{zz}} \left( \mathcal{H}^2 -m^2 g_{tt} g_{x_1 x_1} \right) }\,,
    \label{rEOM}
\end{equation}
Substituting the expressions of various metric functions, we get
\begin{equation}
  \dot{z} = \frac{1}{\mathcal{H}} \sqrt{g_s \left( \mathcal{H}^2 -\frac{m^2  e^{4 A_s (z)}}{z^4} \right)}\,.
    \label{rEOM}
\end{equation}
The above equation immediately implies two turning points for the geodesic, defined by $\dot{z}(t)=0$. One occurs at $z=z_0$, when $g_s(z_0)=0$, while the other is determined by the condition $\mathcal{H} = \frac{m e^{2A_s(z)}}{z^2}$, corresponding to a finite radial position whose value depends on the conserved energy $\mathcal{H}$. In particular, taking a large radial cutoff placed close to the asymptotic AdS boundary at $z=z_{UV}$, fixes the conserved Hamiltonian to
\begin{equation}
\mathcal{H}=\frac{m e^{2A_s(z_{UV})}}{z_{UV}^2}\,.
\end{equation}
To determine the trajectory of the string probe, we again set the integration constant to zero and impose the same initial conditions as in the last section
\begin{equation}
z(0)=z_{UV}=\epsilon, \qquad \dot{z}(0)=0\,.
\label{initialcondstringsoli}
\end{equation}
This corresponds to releasing the probe string from rest in the UV region and allowing it to evolve solely under the influence of the background geometry. With these initial conditions, the conserved Hamiltonian is fixed by its boundary value,
which becomes parametrically large as the cutoff approaches the boundary. In the dual field theory, this initial configuration corresponds to preparing a highly energetic ultraviolet excitation, whose subsequent evolution is holographically encoded in the radial motion of the bulk probe governed by Eq.~(\ref{rEOM}).

Because of the intricate functional forms of $g_s(z)$ and $A_s(z)$, the trajectory cannot be determined analytically in closed form. We therefore resort to a numerical analysis to obtain the corresponding solutions. But before moving on to plot the direct integration of our geodesic, we analyze the behavior of our geodesic at the UV and the IR boundaries. To do so, we rewrite Eq.~(\ref{rEOM}) as
\begin{equation}
    \dot{z}(t)=\sqrt{f(z)g_s}\,, 
    \label{eq:rdot}
\end{equation}
where 
\begin{equation*}
    f(z)=1-\frac{z_{UV}^4 e^{4A_s(z)}}{z^4 e^{4A_s(z_{UV})}}\,.
    \label{eq:rdot}
\end{equation*}
Upon expanding round $z= z_{UV}$, $f(z)$ becomes
\begin{equation}
    f(z\rightarrow z_{UV})\sim 4\left(\frac{1}{z_{UV}}-A_s'(z_{UV})\right)(z-z_{UV}) + \mathcal{O} (z-z_{UV})^2 \,,
\end{equation}
Accordingly, near $z_{UV}$, Eq.~(\ref{rEOM}) becomes
\begin{equation}
    \frac{\partial z}{\partial t}\sim 2\sqrt{\left(\frac{1}{z_{UV}}-A_s'(z_{UV})\right)g_s(z_{UV})(z-z_{UV})} + \mathcal{O} (z-z_{UV})^{\frac{3}{2}} \,.
\end{equation}
This gives us 
\begin{equation}
    z=z_{UV} + g_s(z_{UV})\big(\frac{1}{z_{UV}}-A_s'(z_{UV})\big)t^2+\mathcal{O}(t^4)\,.
    \label{eq:ratUV}
\end{equation}
Explicitly, we have
\begin{equation}
z(t)=z_{UV}
+
\left[
\frac1{z_{UV}}-\sqrt{6a}
+2az_{UV}+\dots
\right]t^2+\mathcal O(t^4).
\label{rUVstringsoliton}
\end{equation}
Eq.~(\ref{eq:ratUV}) gives us the idea of the geodesic behavior in the near UV region. It shows that the probe string initially departs from the UV cutoff quadratically in time.  
This also shows that the leading near-boundary behavior is identical to that of pure AdS, while the effect of the scalar deformation enters only through subleading corrections.

To analyze the geodesic behavior near the geometric cutoff region $z\sim z_0$, we analyze the behavior of $f(z)$ and $g_s(z)$ near the IR cutoff.
\begin{align}
    f(z)={}&1-\frac{z_{UV}^4 e^{4 A_s(z_0)-4 A_s(z_{UV})}}{z_0^4}\\ \nonumber
    &-4\frac{z_{UV}^4}{z_0^4}
e^{4[A_s(z_0)-A_s(z_{UV})]}
\left(\frac1{z_0}-A_s'(z_0)\right)(z_0-z)
+\mathcal O\!\left((z_0-z)^2\right),
\end{align}
and 
\begin{equation}
g_s(z)=
\frac{18a^2z_0^3e^{3az_0^2}}
{e^{3az_0^2}(3az_0^2-1)+1}
(z_0-z)
+\mathcal O\!\left((z_0-z)^2\right),
\end{equation}
which gives us the geodesic equation near the IR cutoff as
\begin{equation}
\frac{dz}{dt}
=
\sqrt{
\frac{18a^2z_0^3e^{3az_0^2}}
{e^{3az_0^2}(3az_0^2-1)+1}
\left[
1-\frac{z_{UV}^4}{z_0^4}
e^{4[A_s(z_0)-A_s(z_{UV})]}
\right](z_0-z)
}
+\mathcal O\!\left((z_0-z)^{3/2}\right).
\end{equation}
After integration, we get
\begin{equation}
    z=z_0-\frac{1}{4} k(z_0) (t_e-t)^2+\mathcal{O}(t_e-t)^4 \,,
    \label{rtIRsoliton}
\end{equation}
where 
\begin{equation}
k(z_0)=
\frac{18a^2z_0^3e^{3az_0^2}}
{e^{3az_0^2}(3az_0^2-1)+1}
\left[
1-\frac{z_{UV}^4e^{4A_s(z_0)}}{z_0^4e^{4A_s(z_{UV})}
}
\right].
\end{equation}
Thus, the string approaches the infrared boundary $z_0$ quadratically in time, with the coefficient determined by the warp factor $A_s(z)$ evaluated at $z=z_0$. Also, the string reaches $z_0$ with zero velocity $\dot{z}(t_e)=0$, while its acceleration remains finite. Thus the trajectory is parabolic in time near $z_0$. Since the radial coordinate also depends quadratically on time near the UV boundary, the UV and IR behaviors are symmetric: the radial velocity vanishes at both endpoints, leading to a quadratic time dependence in both regimes.

\begin{figure*}[htbp]
    \centering
    \begin{subfigure}[b]{0.48\textwidth}
        \centering
\includegraphics[width=\linewidth]{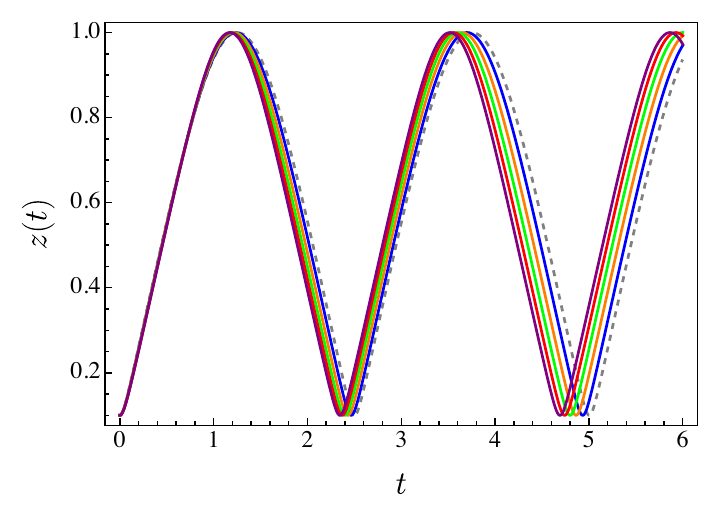}
        \caption{Trajectory $z(t)$ for various values of $a$. Here $z_0 = 1$ and $z_{UV} = 0.1$ are used. The grey dashed, blue, orange, green, red, and purple curves correspond to $a = 0$, $0.1$, $0.2$, $0.3$, $0.4$, and $0.5$, respectively.}
        \label{fig:rvstavary}
    \end{subfigure}
        \hfill
        \begin{subfigure}[b]{0.48\textwidth}
        \includegraphics[width=\linewidth]{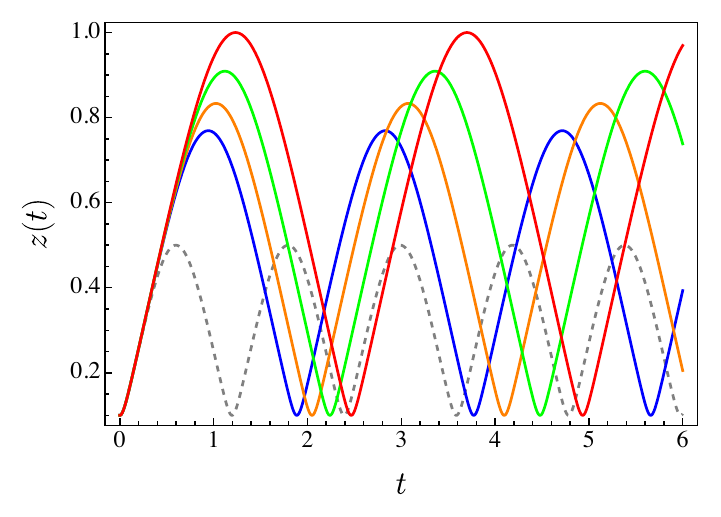}
        \caption{Trajectory $z(t)$ for various values of $z_0$. Here $a = 0.1$ and $z_{UV} = 0.1$ are used. The red, green, brown, blue, and grey dashed curves correspond to $z_0 = 1$, $0.9$, $0.8$, $0.75$, and $0.5$, respectively.}
        \label{fig:rvstr0vary}
    \end{subfigure}
    \caption{The trajectory $z(t)$ of the probe string under varying parameters $a$ and $z_0$.}
    \label{fig:rtVStAdSsolitonstring}
\end{figure*}

The numerically obtained entire evolution of the radial profile of the probe string $z(t)$ is shown in Figure~\ref{fig:rvstavary}. Here we keep the IR cutoff distance fixed at $z_0=1$ and plot the trajectory with varying $a$. Because the motion is naturally bound by the tip of the soliton $z_0$ and the UV cutoff $z_{UV}$, the trajectories exhibit periodic motion, with the oscillation period depending sensitively on the parameter $a$. An interesting feature is that the trajectories nearly coincide during the initial stage of the evolution, while noticeable differences emerge only at later times. In particular, increasing $a$ causes the probe to reach $z_0$ more rapidly. This behavior originates from the exponential warp factor, whose influence becomes increasingly pronounced deeper in the bulk. We also find that although the parameter $a$ modifies the oscillation frequency, the amplitude remains essentially unaffected.

The trajectory for different $z_0$ values is shown in Figure~\ref{fig:rvstr0vary}. As the maximum radial depth is set by $z_0$, the temporal behavior of the trajectories depends strongly on $z_0$. For smaller values of $z_0$, the probe oscillates rapidly between the boundary and $z_0$. As $z_0$ increases, the oscillation period becomes longer, and the corresponding frequency decreases. These results indicate that while $z_0$ fixes the maximum radial extent of the motion, it primarily controls the timescale governing the string dynamics in the bulk.

In the current case, to compute the complexity from the radial proper momentum, we first define a proper coordinate $y$, such that 
\begin{equation}
dy^2=\frac{e^{2A_s(z)}dz^2}{z^2 g_s(z)}\,.
\end{equation}
Then the proper momentum is given by
\begin{equation}
\frac{d C}{dt} \thicksim P_y=P_z\frac{d\dot{z}}{d\dot{y}}\,.
\end{equation}
This leads to
\begin{equation}
\frac{d C}{dt} \thicksim P_y= \frac{m \dot{z}\sqrt{g_{zz}g_{x_1x_1}}}{\sqrt{g_{tt}-g_{zz}\dot{z}^2}}\,,
\end{equation}
Explicitly, for the AdS-soliton, we have
\begin{equation}
P_y= \frac{m \dot{z} e^{A_s(z)}}{z \sqrt{g_s (z) \left(1-\frac{\dot{z}^2}{ g_s(z)}  \right)}} = \frac{z\dot{z} \mathcal{H}}{ e^{A_s(z)} \sqrt{g_s(z)}}\,.
\end{equation}
Using Eqs.~(\ref{eq:ratUV}) and (\ref{rtIRsoliton}), the expansions of $P_y$ close to UV and IR turning point are
\begin{eqnarray}
P_y\bigg\rvert_{t\thicksim 0}   & \propto & \frac{2m e^{A_s(z_{UV})}}{z_{UV}}
\sqrt{g(z_{UV})}
\left(\frac1{z_{UV}}-A_s'(z_{UV})\right)t
+\mathcal O(t^3), \nonumber \\
P_y\bigg\rvert_{t\thicksim t_e}  & \propto & P_0+
\frac{k_0}{4P_0}
\left(A_s'(z_0)-\frac1{z_0}\right)
\left(\frac{\mathcal{H}^2}{U_0}+m^2U_0\right)(t_e-t)^2
+\mathcal O((t_e-t)^4),
\end{eqnarray}
where $U_0=e^{2A_s(z_0)}/z_0^2$
and $P_0=\sqrt{\frac{\mathcal{H}^2}{U_0}-m^2U_0}$.

It is interesting to see that although $z(t)$ is quadratic near both $z_{UV}$ and $z_0$, the proper momentum (and hence the complexity) has distinct time dependence near those points. The proper momentum, in particular, vanishes near $z_{UV}$ due to the boundary condition, whereas it has a finite value near $z_0$. 

\begin{figure*}[htbp]
    \centering
    \begin{subfigure}[b]{0.48\textwidth}
        \centering
        \includegraphics[width=\linewidth]{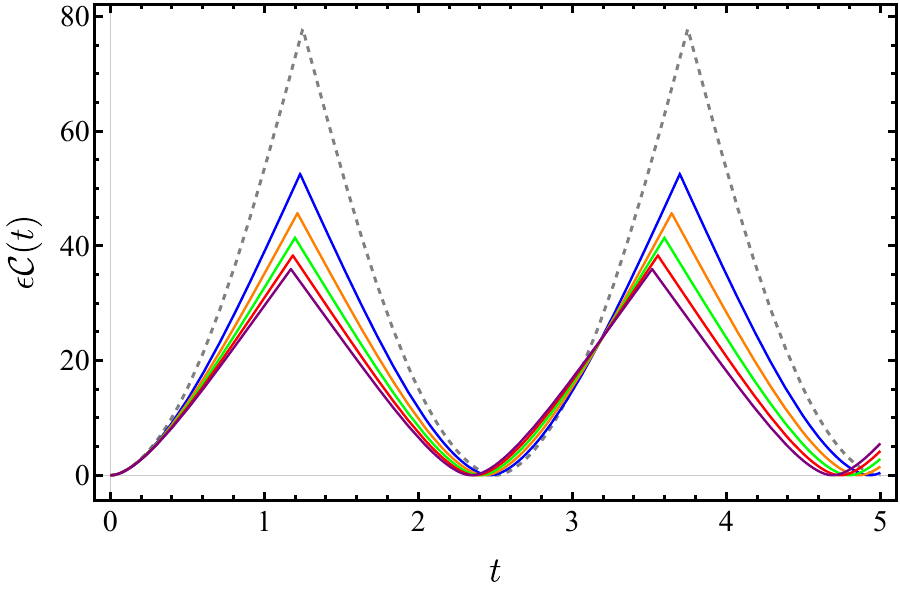}
        \caption{Complexity $C(t)$ for extended operator with various values of $a$.}
    \label{fig:Ctvstavary}
    \end{subfigure}
    \hfill 
    \begin{subfigure}[b]{0.48\textwidth}
        \centering
        \includegraphics[width=\linewidth]{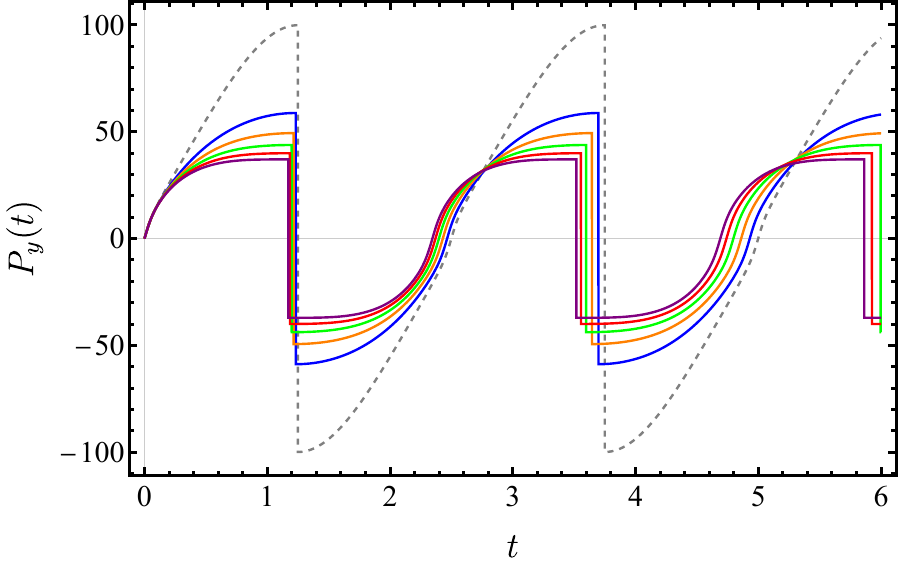}
    \caption{Momentum $P_y(t)$ for extended operator with various values of $a$.}
    \label{fig:Pytvstavary}
    \end{subfigure}
    
    \caption{Complexity and momentum of the probe string under varying parameters $a$. Here $z_0 = 1$ and $z_{UV}=0.1$ are used. The grey dashed, blue, orange, green, red,  and purple curves correspond to 
        $a = 0$, $0.1$, $0.2$, $0.3$, $0.4$, and $0.5$, respectively.}
    \label{fig:CtPytVStsoliton}
\end{figure*}

\begin{figure*}[htbp]
    \centering
    \begin{subfigure}[b]{0.48\textwidth}
        \centering
        \includegraphics[width=\linewidth]{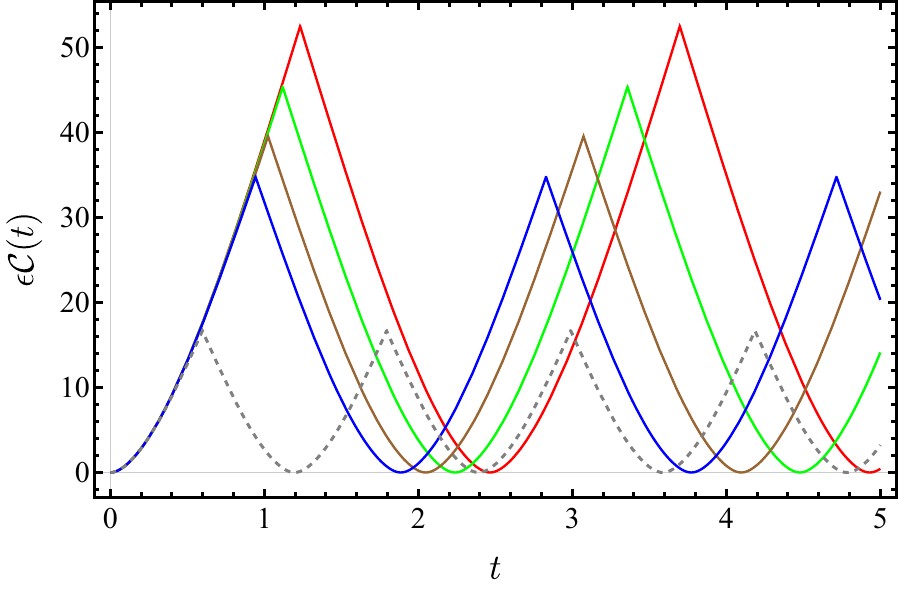}
        \caption{Complexity $C(t)$ for extended operator with various values of $z_0$.}
    \label{fig:Ctvstavary}
    \end{subfigure}
    \hfill 
    \begin{subfigure}[b]{0.48\textwidth}
        \centering
        \includegraphics[width=\linewidth]{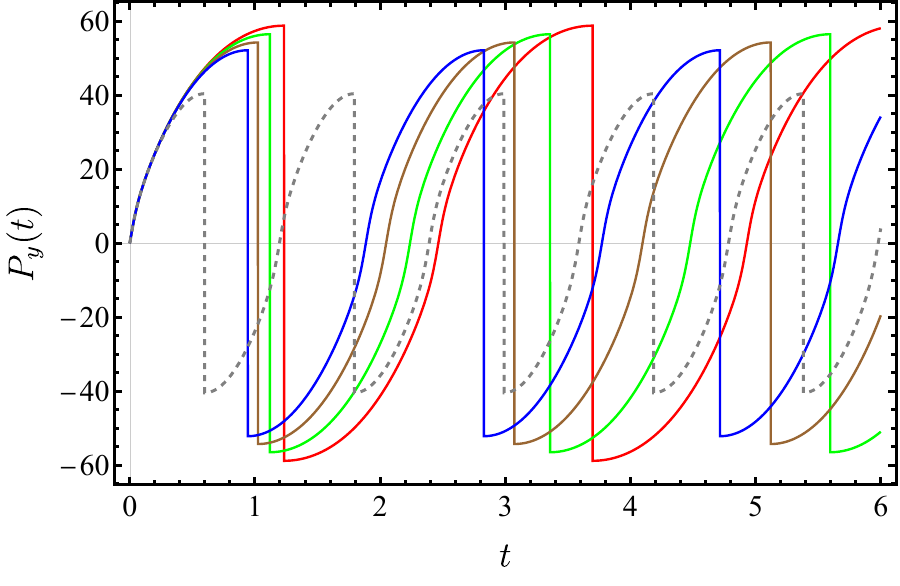}
    \caption{Momentum $P_y(t)$ for extended operator with various values of $z_0$.}
    \label{fig:Pytvstavary}
    \end{subfigure}
    
    \caption{Complexity and momentum of the probe string under varying parameters $z_0$. Here $a = 0.1$ and $z_{UV}=0.1$
are used. The red, green, brown, blue, and grey dashed curves correspond to $z_0 = 1$, $0.9$, $0.8$, $0.75$, and $0.5$, respectively.   }
    \label{fig:CtPytVStr0varysoliton}
\end{figure*}

Figures~\ref{fig:CtPytVStsoliton} and \ref{fig:CtPytVStr0varysoliton} show the time evolution of the Krylov complexity and proper momentum for several values of $a$ and $z_0$. The grey dashed curve in Figure~\ref{fig:CtPytVStsoliton} corresponds to the standard AdS-soliton background and is recovered in the limit $a \rightarrow 0$. We observe that both quantities vary monotonically with time up to $t=t_e$. Beyond this point, the motion reverses, and both quantities retrace their evolution, reflecting the oscillatory nature of the trajectory. The oscillatory pattern originates from the fact that the probe string is confined between the ultraviolet boundary and the infrared end of the geometry, so that its motion is restricted to a finite radial interval. As a consequence, the dual operator evolves within a finite-dimensional effective Hilbert space and repeatedly revisits previously explored regions of the operator space, giving rise to recurrent dynamics. From the boundary-theory perspective, the confinement of the bulk probe prevents quantum information from spreading indefinitely, causing the complexity to undergo repeated growth and return cycles rather than exhibiting unbounded late-time growth \cite{Fatemiabhari:2025usn}. From these figures, the proper momentum may appear to be discontinuous. However, at $t=t_e$, the momentum remains continuous; only its direction changes, albeit very abruptly. We also observe that the time dependence of both the Krylov complexity and the proper momentum near $t=0$ and $t=t_e$ is in agreement with the analytic behavior derived in the corresponding UV and IR regimes. Moreover, the magnitudes of both the complexity and the momentum decrease with increasing $a$.

\subsection{In the deformed thermal-AdS background}
We now turn to the spread complexity of an infalling fundamental string in the deformed thermal-AdS background, which also describes a confining phase of the dual theory. Most of the equations and computations are analogous to the AdS-soliton case. 

A consistent embedding would be of the form
\begin{eqnarray}
t=\tau\,,~~~z=z(t)\,,~~~x_1=x_1(\sigma)\,.
\label{stringembedinAdS}
\end{eqnarray}
On this embedding, the NG action reduces to
\begin{eqnarray}\label{NGaction}
S & = & - m \int dt  \sqrt{\left(g_{tt}-g_{zz}\dot{z}^2 \right)g_{x_1 x_1}} = - m \int dt \mathcal{L}\,,
\end{eqnarray}
where the constant \(m\) is defined in Eq.~(\ref{mexpstring}). The canonical momentum and Hamiltonian are now given by
\begin{eqnarray}
P_z = \frac{m ~ \dot{z} ~ e^{2A_s(z)} }{z^2\sqrt{1-\dot{z}^2}}\,, \qquad
\mathcal{H} = \frac{m  ~ e^{2A_s(z)} }{z^2\sqrt{1-\dot{z}^2}}\,.
\label{conservedmomstrinhther}
\end{eqnarray}
Squaring both sides of (\ref{conservedmomstrinhther}) and solving $\dot{z}$, we get the following first-order equation of motion:
\begin{equation}
  \dot{z} = \frac{1}{\mathcal{H}} \sqrt{\left( \mathcal{H}^2 -\frac{m^2 e^{4 A_s (z)}}{z^4} \right)} = \sqrt{\mathcal{F}}\,,
    \label{rEOMstringthermal}
\end{equation}
The initial conditions (\ref{initialcondstringsoli}) again fixes the conserved Hamiltonian to $\mathcal{H}=\frac{1}{z_{UV}^2}m  e^{2A_s(z_{UV})}$.

\begin{figure*}[htbp]
    \centering
\begin{subfigure}[b]{0.48\textwidth}
        \centering
        \includegraphics[width=\linewidth]{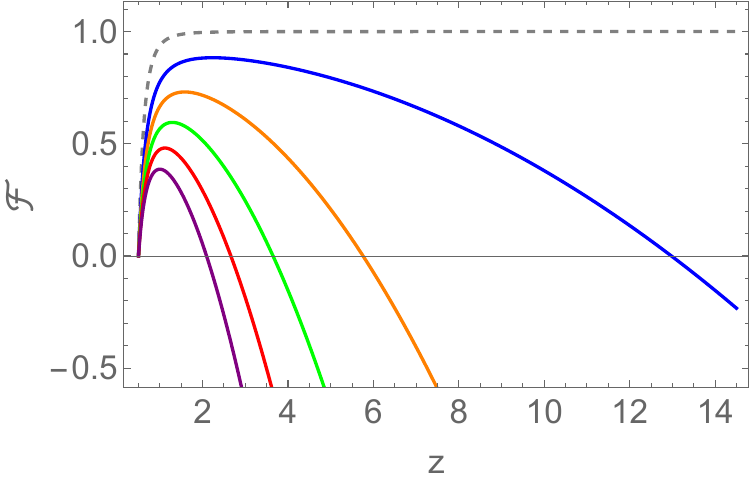}
    \caption{Turning point condition. Here, $z_{UV}=0.5$ is used for better visualization of the IR roots.}
    \label{fig:turningstringAdS}
    \end{subfigure}
     \hfill 
    \begin{subfigure}[b]{0.48\textwidth}
        \centering
        \includegraphics[width=\linewidth]{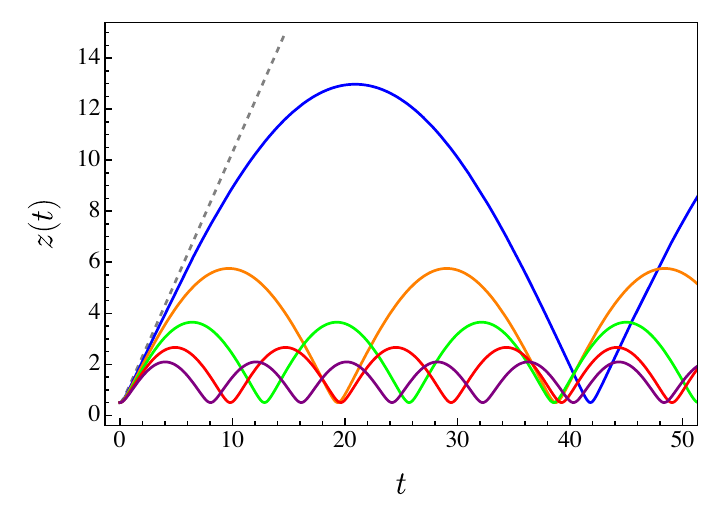}
        \caption{Trajectory $z(t)$ for extended operator with various values of $a$. Here, $z_{UV}=0.5$ is used for illustration.}
    \label{fig:rtvstgs1avary}
    \end{subfigure}
    \caption{Turning point condition and radial motion of the probe string under varying parameters $a$. The blue, orange, green, red, and purple curves correspond to $a=0.1$, $0.2$, $0.3$, $0.4$, and $0.5$, respectively. The grey dashed curve corresponds to $a=0$ (pure AdS). } 
    \label{fig:rtvstAdS}
\end{figure*}

Let us now analyze the turning point condition $\dot{z}=0$ of the above geodesic equation. This is equivalent to requiring \begin{equation}
1 - \frac{z_{UV}^4 e^{4 A_s (z)}}{z^4 e^{4 A_s (z_{UV})}} =0 \,.  
\end{equation}
An interesting feature of the turning-point equation is that, in contrast to the pure AdS geometry, it admits a solution at a finite radial position in the infrared region. The location of this turning point ($z_*$) is shown in Figure~\ref{fig:turningstringAdS}. In the limit $a\to0$, corresponding to the pure AdS background, the turning point disappears, and the string can propagate arbitrarily deep into the bulk. For any nonzero value of $a$, however, the string-frame warp factor produces an effective ``\textit{infrared wall}'' that prevents further radial motion, forcing the string to reverse direction and move back toward the boundary. The resulting trajectory therefore exhibits a bounce in the deep bulk, closely resembling the behavior observed in the AdS-soliton background.

An additional feature is that the turning point moves toward smaller values of $z$ as the parameter $a$ increases. Consequently, the radial region accessible to the string becomes progressively smaller for stronger scalar deformations. Since the string traverses a shorter radial interval before reversing its motion, the time required to reach the turning point decreases with increasing $a$. As a result, the oscillation period becomes shorter, or equivalently, the oscillation frequency increases as $a$ is increased. This trend mirrors the behavior found earlier in the deformed AdS-soliton background, indicating that the scalar deformation plays a similar dynamical role in both confining geometries.

Having analyzed the turning point structure, we now examine the behavior of the trajectory in the ultraviolet and infrared regions separately. The asymptotic analysis in these two regimes provides useful analytic insight into the structure of the string motion and will serve as a guide for interpreting the full numerical solutions presented later. Expanding around $z=z_{UV}$ gives
\begin{equation}
1-\frac{z_{UV}^4e^{4A_s(z)}}{z^4e^{4A_s(z_{ UV})}} = 4\left( \frac{1}{z_{UV}} - A_s'(z_{UV})
\right) (z-z_{UV}) +\mathcal{O}\!\left[(z-z_{UV})^2\right].
\end{equation}
Substituting this into Eq.~(\ref{rEOMstringthermal}) and approximating $z\simeq z_{UV}$ in the prefactor, one obtains
\begin{equation}
\frac{dz}{dt} = 2 \sqrt{
\left(\frac{1}{z_{UV}}-A_s'(z_{UV})\right)
(z-z_{\rm UV})} + O[(z-z_{\rm UV})^{3/2}].
\end{equation}
Integrating with the initial conditions yields
\begin{equation}
z(t)=z_{UV} + \left(\frac{1}{z_{UV}}-A_s'(z_{UV})\right)t^2 +O(t^4)\,.
\end{equation}
Therefore, the radial trajectory at early times is given by
\begin{equation}
z(t) = z_{UV} + \left(\frac{1}{z_{UV}} + 2az_{UV} - \sqrt{2a(2az_{UV}^2 +3)} \right)t^2
+\mathcal{O}(t^4).
\label{rUVstringAdS}
\end{equation}
Again the probe string departs from the UV cutoff quadratically in time. This is similar to the deformed AdS-soliton case discussed earlier. In fact, the coefficient of the $t^2$ term is identical in both cases. This agreement is expected because the asymptotic structure of the deformed thermal-AdS and deformed AdS-soliton geometries coincides near the boundary to leading order.

Similarly, to determine the behavior of the trajectory near the infrared turning point $z=z_*$, we consider $z=z_*-\delta z$, with $\delta z\ll z_*$, and expand the equation of motion around $z=z_*$. This gives
\begin{equation}
1-\frac{z_{\rm UV}^4e^{4A_s(z)}}{z^4e^{4A_s(z_{\rm UV})}} = \left(4A_s'(z_*)-\frac{4}{z_*}\right)(z_*-z)
+\mathcal{O}(z_*-z)^2.
\end{equation}
Accordingly, the equation of motion reduces to
\begin{equation}
\frac{dz}{dt} =2\sqrt{\left(A_s'(z_*)-\frac{1}{z_*}\right)(z_*-z)} + \mathcal{O}(z_*-z)^{3/2}\,.
\end{equation}
Integrating with the condition $z(t_e)=z_*$, with $t_e$ being the time required to reach the turning point, gives
\begin{equation}
t_e - t = \frac{\sqrt{z_*-z}}{\sqrt{\left(A_s'(z_*)-1/z_*\right)}} +  \mathcal{O}(z_*-z)^{3/2},
\end{equation}
Therefore,
\begin{equation}
z(t) = z_* - \left(
A_s'(z_*)-\frac{1}{z_*} \right) (t_e-t)^2 + \mathcal{O}(t_e-t)^4.
\label{rIRstringAdS}
\end{equation}
Therefore, as in the deformed AdS-soliton background, the string approaches the infrared turning point quadratically in time, with the coefficient governed by the local derivative of the warp factor, $A_s'(z_*)$. Unlike the deformed AdS-soliton case, however, the coefficient of the $(t_e-t)^2$ term is different, reflecting the distinct bulk geometries of the two confining backgrounds.

The full numerical evolution of the radial profile of the probe string, $z(t)$, is shown in Figure~\ref{fig:rtvstgs1avary}. Here, the same color scheme as in Figure~\ref{fig:rtVStAdSsolitonstring} is used for ease of comparison. As in the deformed AdS-soliton background, the trajectories exhibit periodic oscillations, with the oscillation period depending sensitively on the deformation parameter $a$. The curves nearly overlap during the early stages of the evolution and begin to separate only at later times. Moreover, the time $t_e$ required to reach the turning point decreases with $a$, which is quite similar to the behavior observed in the deformed AdS-soliton case.

The proper momentum and the corresponding Krylov complexity are now given by 
\begin{equation}
\frac{d C}{dt} \thicksim P_y = \frac{m \dot{z} e^{A_s(z)}}{z \sqrt{\left(1-\dot{z}^2  \right)}} = \frac{\dot{z}z \mathcal{H}}{ e^{A_s(z)} }\,,
\end{equation}
where $P_y$ is a proper momentum conjugate to the proper radial coordinate $y$, defined through $dy=e^{A_s(z)}dz/z$. Using Eqs.~(\ref{rUVstringAdS}) and (\ref{rIRstringAdS}), the  expansions of $P_y$ close to UV and IR turning point are
\begin{eqnarray}
    P_y\bigg\rvert_{t\thicksim 0}  & \propto & \frac{2me^{A_s(z_{\rm UV})}}{z_{\rm UV}}\big(\frac{1}{z_{\rm UV}}-A_s'(z_{\rm UV})\big)t+\mathcal{O}(t^3), \nonumber \\
    P_y\bigg\rvert_{t\thicksim t_e} & \propto & \frac{2mz_*e^{2A_s(z_{\rm UV})}}{z_{\rm UV}^2 e^{A_s(z_*)}}\big(A_s'(z_*)-\frac{1}{z_*} \big)(t_e-t)+\mathcal{O}(t_e-t)^3
\end{eqnarray}
Therefore, in both the UV and IR regimes, the momentum varies linearly with time, implying a quadratic time dependence of the complexity near the endpoints of the motion, with the slope receiving corrections from the scalar deformation parameter $a$. The UV behavior is quite similar to the soliton case. However, there are drastic differences in their IR behavior. In particular, $P_y$ exhibits a linear time dependence in the IR and vanishes at the IR turning point. By contrast, in the AdS-soliton background, $P_y$ remained finite at the IR turning point $z_0$ and exhibited a quadratic time dependence near that point.

\begin{figure*}[htbp]
    \centering
    \begin{subfigure}[b]{0.48\textwidth}
        \centering
        \includegraphics[width=\linewidth]{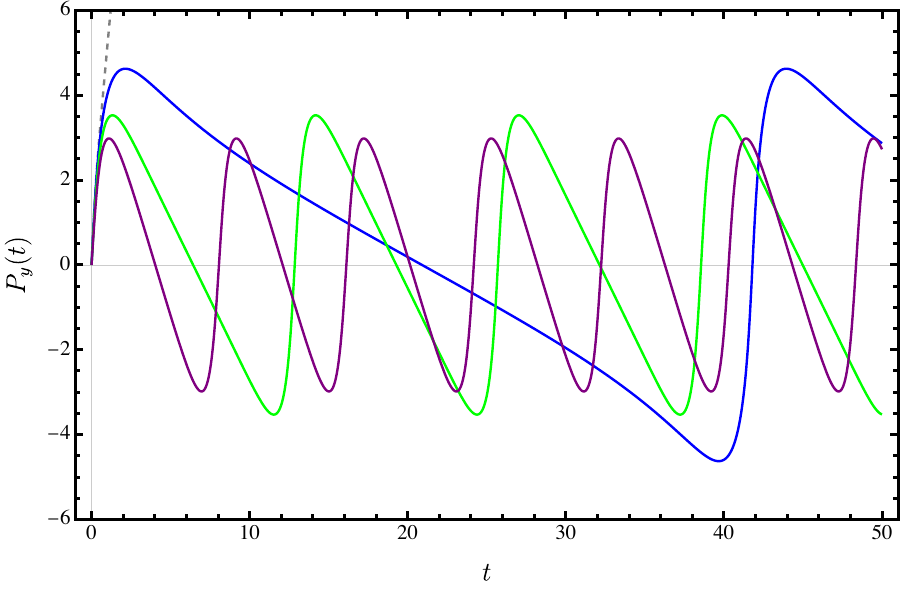}
        \caption{Proper momentum $\mathcal{P}_y$ for the string probe with various values of $a$.}
    \label{fig:pyvstgs1avary}
    \end{subfigure}
    \hfill 
    \begin{subfigure}[b]{0.48\textwidth}
        \centering
        \includegraphics[width=\linewidth]{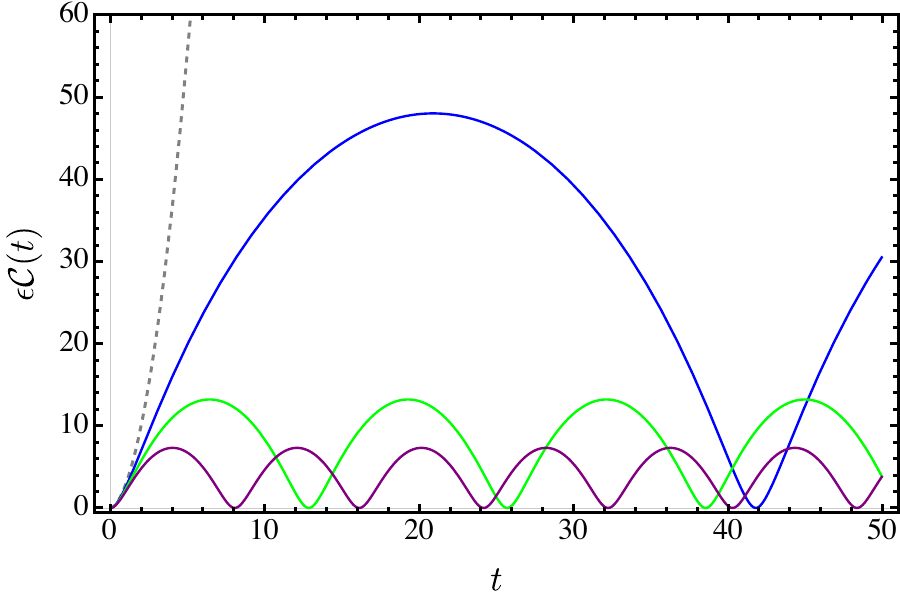}
    \caption{Complexity $C(t)$ for the string probe with various values of $a$.}
    \label{fig:Ctvstavary}
    \end{subfigure}
    
    \caption{Proper momentum and complexity of the probe string under varying parameters $a$. Here $z_{UV}=0.5$ is used. The blue, green, and purple curves correspond to $0.1$, $0.3$,  and $0.5$, respectively. The grey dashed curve corresponds to $a=0$ (pure AdS).}
    \label{fig:CtPytVSt}
\end{figure*}

Figures~\ref{fig:pyvstgs1avary} and \ref{fig:Ctvstavary} illustrate the time evolution of the proper momentum $P_y(t)$ and the complexity $C(t)$ associated with the infalling string probe for various values of the deformation parameter $a$. We observe that the oscillation frequency scales with $a$, while the peak proper momentum scales inversely with $a$. The observed oscillatory behavior arises from the infrared wall identified in Figure~\ref{fig:turningstringAdS}, which reflects the confining nature of the corresponding dual boundary theory, analogous to the soliton case. Despite this similarity in confinement, it is crucial to emphasize the distinct dynamical behavior of the string probe in the deformed thermal-AdS background compared to the soliton background. In the soliton case, the proper momentum increases monotonically to a maximum before experiencing a sign reversal. In the deformed thermal-AdS case, however, $P_y(t)$ exhibits a smooth nonmonotonic transient profile: it initially increases from zero to an intermediate maximum, subsequently decreases, vanishes at the turning point $t=t_e$, and only then smoothly reverses sign. Thus, unlike the soliton case, where the momentum increases continuously before changing direction, the deformed thermal-AdS background displays a distinct deceleration stage in which the momentum is continuously driven to zero before the reversal occurs. This difference in momentum dynamics is directly reflected in the holographic complexity, which can be geometrically interpreted as the area under the $P_y(t)$ curve. Consequently, the complexity associated with the string probe oscillates between the UV boundary and the effective IR wall. Moreover, the complexity curve exhibits a smooth turnaround in the deep IR region. This smoothness, which is distinctly absent in the soliton case, is a direct physical consequence of the proper momentum smoothly vanishing at the turning point, i.e., $P_y(t\rightarrow t_e)=0$.

\section{Back to the spread complexity for an infalling probe particle, but now in the string frame}\label{sec:particleStrframe}
In the last section, we saw that the Krylov complexity associated with a probe string exhibits oscillatory behavior in the confined phase dual to the deformed AdS-soliton and the deformed thermal-AdS. In the deformed AdS-soliton, the reason for this oscillatory behavior was straightforward, as the geometry has a natural cut-off both in the UV and in the IR. What was more surprising was the fact that oscillatory behavior also appeared in the deformed thermal-AdS background, even though there is no genuine IR cutoff in the geometry. In that case however, the oscillation originates from the appearance of an effective ``imaginary wall'' beyond which the string cannot propagate. Note that the computations in the last section were done in the string frame, as the dynamics of a fundamental string is naturally described in the string frame, see e.g.~\cite{Alvarez:2001qj} for a discussion about the string vs.~Einstein frame.

The latter frame is more natural from the viewpoint of gravitational dynamics of point particles. As such, in Section~\ref{sec:pSoliton}, the Krylov complexity associated with a probe particle was computed in the deformed AdS-soliton and deformed thermal-AdS in the Einstein frame. It was observed that the complexity exhibits exponential growth in the confined phase dual to the deformed thermal-AdS background, as opposed to the oscillatory behavior in the deformed AdS-soliton background. This difference is rather striking because, after all, both geometries describe the confining phase and contrast with the behavior of other information-theoretic observables, such as entanglement entropy, which displays qualitatively similar features in the two backgrounds.  This naturally raises the question of whether the particle complexity in the deformed thermal-AdS geometry can also be made oscillatory. Taking hints from the string dynamics, one possible way to do that is by working in the string frame. Although not reported here, we have verified that, if the same analysis is performed in the Einstein frame, the oscillatory behavior disappears and is replaced by exponential growth in the confined phase dual to the deformed thermal-AdS.

Therefore, we will now evaluate the probe particle dynamics and associated complexity also in the string frame. As we will soon reveal, the string-frame geometry generates a turning point in the deep IR region, forcing the particle to reverse its radial motion. Consequently, the particle oscillates between the UV and IR regions, and the associated Krylov complexity exhibits oscillatory behavior in the deformed thermal-AdS background as well.

Since the probe-particle complexity in the deformed AdS-soliton background exhibits qualitatively similar behavior in both the string and Einstein frames, we omit the AdS-soliton analysis for brevity here. Instead, we focus on the probe-particle dynamics in the string frame in the deformed thermal-AdS background, where the frame dependence leads to qualitatively new behavior. The dynamics of the particle is governed by $z(t)$. Using Eq.~\eqref{thermalAdSback}, the induced line element formulated in the string frame for the infalling particle simplifies to:
\begin{eqnarray}
    ds_{ind}^2 = \frac{e^{2A_s(z)}}{z^2} \left( -1 +\dot{z}^2 \right) dt^2\,.
    \label{eq:metricparticleAdS}
\end{eqnarray}
The action of the falling point particle is\footnote{Notice that \eqref{falling2} differs from \eqref{falling1} as the underlying metrics are different due to the dilaton.}
\begin{equation}\label{falling2}
    \mathcal{S}_P = -m \int dt \mathcal{L}_P, \quad \mathcal{L}_P = \sqrt{-g_{ind}},
\end{equation}
where
\begin{equation}
    \mathcal{L}_P = \frac{e^{A_s(z)}}{z} \sqrt{1  - \dot{z}^2}.
\end{equation}
The canonical momentum and the conserved Hamiltonian are now
\begin{eqnarray}
    P_z & = & \frac{m e^{A_s(z)} \dot{z}}{z \sqrt{1-\dot{z}^2 } }\,, \qquad
    \mathcal{H}  =  \frac{m e^{A_s(z)} }{z\sqrt{1-\dot{z}^2 }}\,,
    \label{eq:HamiltonianparticleAdS}
\end{eqnarray}
This gives the following radial equation of motion:
\begin{equation}
    \dot{z}(t) = \frac{1}{\mathcal{H}} \sqrt{ \mathcal{H}^2 -\frac{ m^2 e^{2A_s(z)}}{z^2} }\,. 
    \label{eq:EOM}
\end{equation}
The initial condition $\dot{z}(0)=0$ and $z(0)=z_{UV}$ fixes the Hamiltonian to $\mathcal{H}=\frac{1}{z_{\rm UV}} m e^{A_s(z_{UV})}$.

\begin{figure*}[htbp]
    \centering
\begin{subfigure}[b]{0.48\textwidth}
        \centering
        \includegraphics[width=\linewidth]{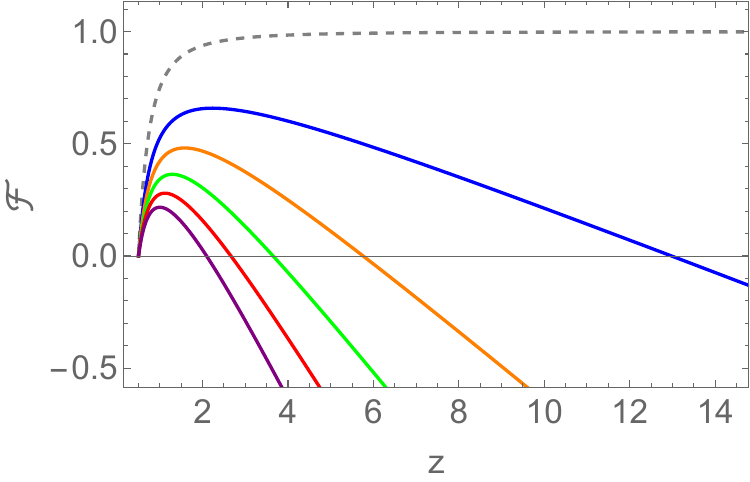}
    \caption{Turning point condition of the probe particle. Here, $z_{UV}=0.5$ is used for a better visualization of the IR roots.}
    \label{fig:turningpointpartcleAdS}
    \end{subfigure}
     \hfill 
    \begin{subfigure}[b]{0.48\textwidth}
        \centering
        \includegraphics[width=\linewidth]{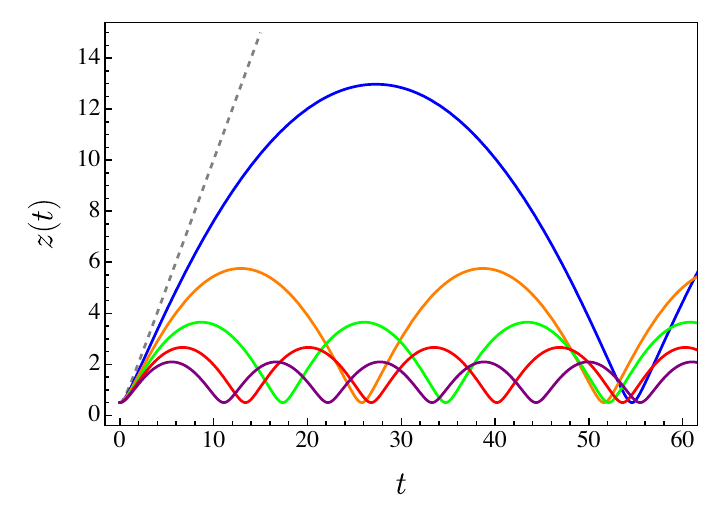}
        \caption{Trajectory $z(t)$ of a probe particle with various values of $a$. Here, $z_{UV}=0.5$ is used for a better visualization.}
    \label{fig:rtvstgs1avaryparticle}
    \end{subfigure}
    \caption{Turning point condition and radial motion of the probe particle in the string frame under varying parameters $a$. The blue, orange, green, red, and purple curves correspond to $a=0.1$, $0.2$, $0.3$, $0.4$, and $0.5$, respectively. The grey dashed curve corresponds to $a=0$ (pure AdS). } 
    \label{fig:rtvstAdSparticle}
\end{figure*}

We now examine the infrared turning point of the particle trajectory by imposing the condition \mbox{$\dot{z}=0$} in the geodesic equation. As in the probe-string analysis, this equation possesses a solution at a finite radial distance in the infrared region. The corresponding turning point, denoted by $z_*$, is displayed in Figure~\ref{fig:turningpointpartcleAdS}. As expected, in the undeformed limit $a\rightarrow0$, no such solution exists, thence allowing the particle to penetrate arbitrarily deep into the bulk. Once the deformation parameter is switched on, however, the string-frame warp factor generates an effective infrared barrier that obstructs further radial motion and causes the particle to turn around and propagate back toward the boundary. Consequently, the particle undergoes a bounce in the deep interior, closely analogous to the behavior found in the AdS-soliton geometry. It is important to emphasize that this effective infrared wall is a feature of the string-frame description and is absent in the Einstein frame.

The leading-order behavior of $z(t)$ at the UV and IR turning point $z_*$ is
\begin{eqnarray}
& & z(t)\bigg\rvert_{t\thicksim 0} =z_{\rm UV}+\frac{1}{2}\bigg(\frac{1}{z_{\rm UV}}-A_s'(z_{\rm UV})\bigg)t^2+\mathcal{O}(t^4)\, , \nonumber \\
& & z(t)\bigg\rvert_{t\thicksim t_e} = z_*-\frac{1}{2}\bigg(A_s'(z_*)-\frac{1}{z_*}\bigg)(t_e-t)^2+\mathcal{O}(t_e-t)^4\,.
\label{ztUVIRpartcilestring}
\end{eqnarray}

Therefore, the trajectory is quadratic with quartic corrections, while the velocity is linear with cubic corrections, in both the IR and UV regions.

The full radial profile of the probe particle is shown in Figure~\ref{fig:rtvstgs1avaryparticle}. The behavior is qualitatively similar to the probe string case in the deformed thermal-AdS background. In particular, the trajectories exhibit periodic oscillations, with the oscillation period depending sensitively on the parameter $a$. The curves nearly overlap during the early stages of the evolution and begin to separate only at later times. Similarly, the time $t_e$ required to reach the turning point decreases with $a$. Consequently, the oscillation frequency increases as the deformation parameter is increased.

The proper momentum and the corresponding Krylov complexity are now given by 
\begin{equation}
\frac{d C}{dt} \thicksim P_y = \frac{m \dot{z}}{\sqrt{1-\dot{z}^2 }} = \frac{z\dot{z} \mathcal{H}}{ e^{A_s(z)} }\,.
\end{equation}
Using Eq.~(\ref{ztUVIRpartcilestring}), the  expansions of $P_y$ close to UV and IR turning point are
\begin{eqnarray}
    P_y\bigg\rvert_{t\thicksim 0}  & \propto & m\bigg(\frac{1}{z_{UV}}-A_s'(z_{UV})\bigg)t+\mathcal{O}(t^3) \,, \nonumber \\
    P_y\bigg\rvert_{t\thicksim t_e} & \propto & m\bigg(A_s'(z_*)-\frac{1}{z_*}\bigg)(t_e-t)+\mathcal{O}(t_e-t)^3
\end{eqnarray}
Hence, both near the UV boundary and near the IR turning point, $P_y$ varies linearly with time. This asymptotic structure is very similar to the case of the probe string in the deformed thermal-AdS background. Importantly, the proper momentum again goes to zero at the IR turning point. 
\begin{figure*}[htbp]
    \centering
\begin{subfigure}[b]{0.48\textwidth}
        \centering
        \includegraphics[width=\linewidth]{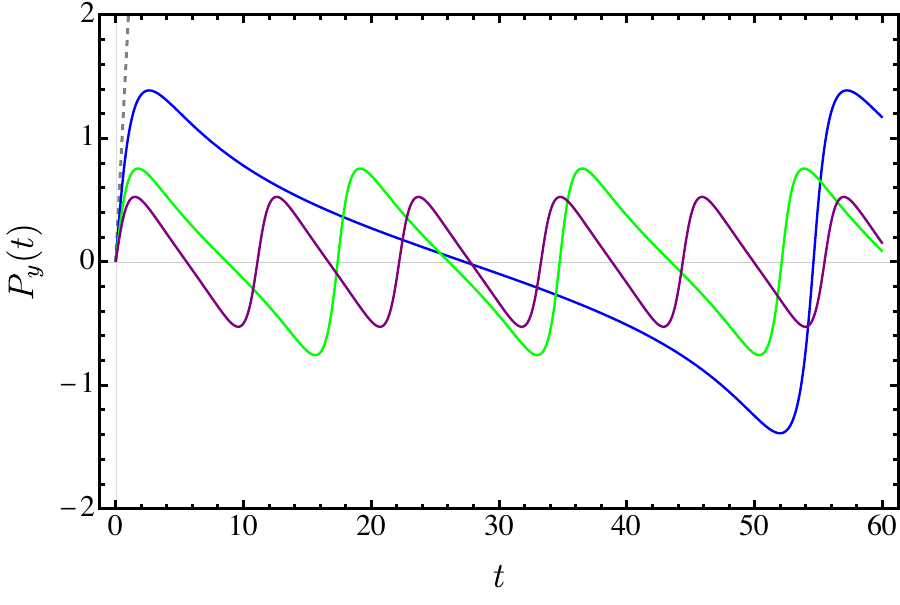}
    \caption{Momentum $P_y(t)$ for probe particle. Here, $z_{UV}=0.5$ is used.}
    \label{fig:Pyvstgs1avaryparticleS}
    \end{subfigure}
     \hfill 
    \begin{subfigure}[b]{0.48\textwidth}
        \centering
        \raisebox{0cm}{\includegraphics[width=\linewidth]{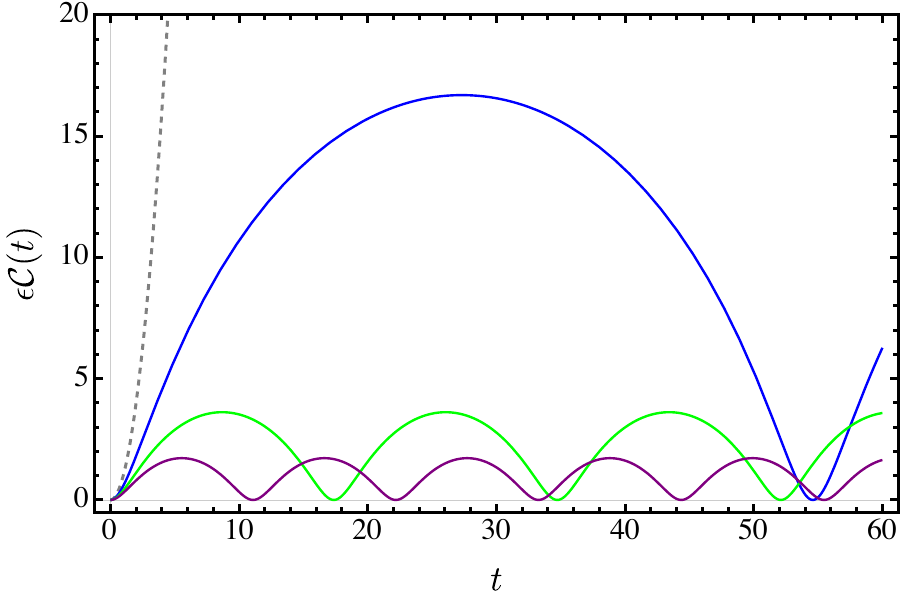}}
        \caption{Associated complexity $C(t)$ of a probe particle. Here, $z_{UV}=0.5$ is used.}
    \label{fig:Ctvstgs1avaryparticle}
    \end{subfigure}
    \caption{Proper momentum and complexity of the probe particle in the string frame under varying parameters $a$. The blue,  green, and purple curves correspond to $a=0.1$,  $0.3$, and $0.5$, respectively. The grey dashed curve corresponds to $a=0$ (pure AdS). } 
    \label{fig:CtvstD0}
\end{figure*}

Figures~\ref{fig:Pyvstgs1avaryparticleS} and \ref{fig:Ctvstgs1avaryparticle} illustrate the time evolution of the proper momentum $P_y(t)$ and the complexity $C(t)$ associated with the infalling point-particle probe for various values of the deformation parameter $a$. The behavior closely resembles that of an infalling string in the deformed thermal-AdS background. In particular, the proper momentum exhibits a smooth nonmonotonic time dependence: it initially increases from zero, reaches a maximum value, subsequently decreases, and vanishes at the turning point $t=t_e$, where it smoothly reverses sign and the direction of motion changes. The complexity is also smooth and increases monotonically until $t=t_e$ after which it begins to decrease monotonically during the outward trajectory.

\section{Conclusion and future perspectives}
\label{Conclusion}
We have investigated holographic Krylov complexity in bottom-up holographic models describing the confined phase of QCD. We employed the simplest Einstein-scalar framework and considered two distinct asymptotically AdS confining geometries, namely the deformed AdS-soliton and deformed thermal-AdS backgrounds. Following the recent proposal of \cite{Caputa:2024sux} on how to compute the information spreading from a holographic perspective, we studied the dynamics of massive particles and strings and analyzed the resulting behavior of the Krylov complexity.

For massive particles described in the Einstein frame, we found a clear distinction between the two confining geometries. In the deformed AdS-soliton background, the radial dynamics is characterized by a turning point and leads to periodic oscillations of the Krylov complexity. Both the period and amplitude of the oscillations depend on the parameters of the holographic background. This behavior is qualitatively consistent with that observed in corresponding top-down confining constructions. In contrast, the deformed thermal-AdS background admits an unbounded radial motion, and the corresponding Krylov complexity grows monotonically with time. Although both geometries describe confined phases of the dual gauge theory, the behavior of Krylov complexity is qualitatively different, indicating that it is sensitive to the detailed infrared structure of the bulk spacetime.

We then extended the analysis to strings, for which the string-frame metric provides the natural description. In this case, the two confining geometries exhibit a remarkably similar qualitative behavior: the Krylov complexity develops an oscillatory structure in both backgrounds. Our analysis indicates that this behavior is associated with an additional turning point generated in the string-frame geometry in the deep infrared. This turning point is not present in the corresponding Einstein-frame description and therefore provides a geometric explanation for the qualitative difference between particle and string dynamics.

The importance of the frame choice becomes even more apparent when massive particle dynamics is reconsidered in the string frame. We find that the particle Krylov complexity in this frame exhibits qualitatively similar behavior in the deformed thermal-AdS and deformed AdS-soliton backgrounds. In particular, the distinction between the oscillatory and nonoscillatory behavior observed for particles in the Einstein frame is substantially modified once the particle dynamics is governed by the string-frame geometry. Essentially, the string frame geometry again develops an additional turning point in the deep IR, leading to oscillatory behavior. These results demonstrate that the qualitative features of holographic information spreading are sensitive not only to the choice of the confined background but also to the frame in which the probe dynamics is formulated. That the frame choice can affect the physics of bottom-up QCD models was already observed in the holographic chaos study of \cite{Shukla:2024qlf}. It is unclear whether this is an unphysical relic of the used bottom-up constructions and whether such differences in predictions would be alleviated in a full-blown top-down setting. In any case, it is in general known that frame choices can be a delicate matter, see ~\cite{Alvarez:2001qj,faraoni1999einstein, cho1992reinterpretation, sk2017scalar, capozziello2010physical, corda2011gravitational, quiros2013conformal, macias2001jordan, galaverni2022jordan, nojiri2001string, jarv2007scalar} for a variety of examples.

The present work opens several directions for future investigation. Firstly, thinking about the just discussed differences between Einstein and string frame results in bottom-up models, it might be instructive to find out if spread complexity could also be studied in a top-down holographic QCD setting, such as \cite{Sakai:2004cn,Sakai:2005yt}\footnote{\cite{Fatemiabhari:2025usn} is already such an attempt in a confining gauge theory, albeit one even somewhat more distant from QCD than the Sakai-Sugimoto construct.}. 

Secondly, to the best of our knowledge, the holographic prescription to connect  the spread complexity rate to the proper momentum of an infalling test particle, is not proven in general, except for the analytically controllable AdS${}_3$/CFT${}_2$ case scrutinized in \cite{Caputa:2024sux}. As such, it would be interesting to study the complexity, and test its holographic interpretation, directly from the Lanczos coefficients in the Hilbert space, although this might be a hard numerical exercise for holographic QCD models.

Furthermore, another extension would be to study probe branes, thereby exploring the complexity associated with operators whose dynamics lie well-beyond the point-particle approximation. Another possible extension would be to study Krylov complexity in a finite-temperature deconfining background, where the thermal effects are expected to modify complexity dynamics. Finally, extending the present analysis to more involved holographic QCD settings, such as with finite chemical potential or in the presence of external magnetic fields, may provide further insight into the behavior of complexity in strongly coupled gauge theories.

\subsection*{Acknowledgments}
J.N. would like to thank Bartosz Pyszkowski for helpful discussions. The work of S.M.~is supported by the core research grant from the Science and Engineering Research Board (now Anusandhan National Research Foundation), a statutory body under the Department of Science and Technology, Government of India, under grant agreement number CRG/2023/007670. The work of D.D.~ was supported by KULeuven IF project~C14/26/108.


\bibliographystyle{JHEP}
\bibliography{biblio}
\end{document}